\documentclass[prc,showpacs,twocolumn,floatfix]{revtex4-1}
\usepackage{graphicx,subfigure}
\usepackage{comment}
\usepackage{amsmath, amsthm, amssymb}
\usepackage{bm}

\def\nuc#1#2{\relax\ifmmode{}^{#1}{\protect{#2}}\else${}^{#1}$#2\fi}

\newcommand {\la} {\langle}\newcommand {\ra} {\rangle}
\newcommand {\beq} {\begin{eqnarray}}
\newcommand {\eeq} {\end{eqnarray}}
\newcommand {\eeqn} [1] {\label{#1} \end{eqnarray}}%
\newcommand {\eol} {\nonumber \\}
\newcommand {\ve} [1] {\mbox{\boldmath $#1$}}

\begin{document}
\graphicspath{{figures/}}

\title{Adiabatic model of $(d,p)$ reactions with expicitly energy-dependent non-local potentials.}
 \author{R.C. Johnson and N.K. Timofeyuk}
 \affiliation{
  Department of Physics,
 Faculty of Engineering and Physical Sciences,
 University of Surrey,
 Guildford, Surrey GU2 7XH, United Kingdom}
\date{\today}

\begin{abstract}

We have developed an approximate way of   dealing with explicit energy-dependence of non-local 
nucleon optical potentials as used to predict the $(d,p)$ cross sections within the
adiabatic theory. Within this approximation, the non-local optical potentials have
to be evaluated at an energy shifted from 
half the incident deuteron energy by the $n-p$ kinetic energy averaged over
 the range of the $n-p$ interaction and then treated as an energy-independent non-local potential.
Thus the evaluation of the distorting    potential in the incident channel is reduced to a problem
solved in our previous work in [{\it Phys. Rev. Lett. 110, 112501(2013) and Phys. Rev. C 87, 064610 (2013)}]. We have demonstrated how our new model works for the case of $^{16}$O$(d,p)^{17}$O,
$^{36}$Ar($d,p)^{37}$Ar and $^{40}$Ca$(d,p)^{41}$Ca reactions and highlighted the need for a
detailed understanding of energy-dependence of non-local potentials. We have also suggested a simple way of correcting the $d-A$ effective potentials for non-locality when the
underlying energy-dependent non-local nucleon potentials are unknown but  energy-dependent
local phenomenological nucleon potentials are available.

\end{abstract}
\maketitle

\section{Introduction}

In our two previous publications \cite{Tim13a,Tim13b} we have proposed a method for calculating
 $A(d,p)B$ cross sections when the interaction of the neutron and proton in deuteron with
the target $A$ is energy-independent but non-local. This development has been motivated by the growing
number of $(d,p)$ experiments performed at radioactive beams facilities with the aim
of extracting spectroscopic information beyond the limits of $\beta$-stability, including cases 
important for astrophysical applications. The $(d,p)$ reaction is known \cite{PTJT} to be dominated
by those components of the total wave functions in which the separation between the neutron 
and proton in the incoming deuteron is less than the range of the $n-p$ interaction. 
Such components are often calculated in the adiabatic distorted wave approximation (ADWA) where
the effective deuteron potential is derived from neutron and proton optical potentials
taken at some fixed energy \cite{JS,HJ}. Since the introduction of ADWA the 
fixed  
neutron and proton energies
used to calculate the adabatic deuteron potential were always taken
to be precisely half of the deuteron incident energy \cite{JS,JT, Joh05}. Such a prescription
seemed to be reasonable from the intuitive point of view that appeals to the low
binding energy of deuteron. 

The first indications for a need to go beyond the approximation of fixed nucleon
energies came from the Faddeev study of transfer reactions in Refs. \cite{Del09a,Del09b}.
The energy dependence of the nucleon optical potentials
 assumed there were  either explicit \cite{Del09a} or
 a result of the non-locality of energy-independent nucleon potential \cite{Del09b}.
Given that the Faddeev formalism is too complicated for the routine analysis of deuteron  stripping
reactions we have suggested \cite{Tim13a,Tim13b} a practical way to account for the non-locality of
nucleon optical potentials in the ADWA developed in \cite{JS,JT}. It turned out  \cite{Tim13b} that for
energy-independent non-local potentials of the Perey-Buck type \cite{PB}
a simple adiabatic deuteron
potential can be constructed as a solution of a transcendental equation, 
similar to the one obtained for nucleon scattering in \cite{PB}. Moreover,
for  $Z = N$ nuclei and isospin independent  nucleon optical  potentials this solution 
is  related to the sum of nucleon potentials taken at an energy shifted
with respect to the half the incident deuteron energy by about 40 MeV. We have shown  that this
large shift comes from the relative $n-p$ kinetic energy averaged over the short-range
potential $V_{np}$ \cite{Tim13a,Tim13b}. The new effective deuteron potential is shallower than that
 traditionally used in the ADWA and this leads to 
a change in normalizaton of predicted $(d,p)$ cross sections with consequences for
their interpretation in terms of nuclear structure quantities.

The simple prescription of Refs. \cite{Tim13a,Tim13b}
to obtain the $d-A$ potential  for $(d,p)$ reactions
relies strongly on the assumption that the energy-dependence of nucleon optical potentials
is a consequence of the non-locality of energy-independent potentials. 
However, the formal theory of
optical potentials developed by Feshbach \cite{Fes58} shows that optical potentials
are not only non-local but energy-dependent as well.  
The explicit energy dependence of Feshbach potentials 
comes from the coupling of channels with the target, $A$, in its ground state to   
channels in which $A$ is excited, i.e, the mechanism that gives rise to the imaginary part of the optical potential when the excited channels are open.
The result is  that global local energy-dependent potentials derived from analysis of  experimental data 
may not be of the Perey-Buck type. We have checked this for  optical potentials taken
from frequently-used global systematics CH89 \cite{CH89}. We found that only
their real parts satisfy the Perey-Buck form with the usual non-locality range
of 0.85 fm while the imaginary parts are definitely not of the Perey-Buck type
and do not follow any immediately obvious systematic rule. Therefore, it is unclear
how to treat explicit energy dependence of the popular CH89 global potential in the ADWA.
This may be relevant to other energy-dependent global optical potentials available in the literature and
used in $(d,p)$ reaction calculations.

In this paper we suggest an approximate practical way of dealing with explicitly energy-dependent non-local optical 
potentials. 
For this purpose we first consider in Sec. II
the connection between three-body model and underlying many-body problem that leads to the
energy dependence of optical potentials. Then in Sec. III we give formal expressions
for optical potential operators in two- and three-body systems. We derive in Sec. IV the 
link between the adiabatic $(d,p)$ models with energy-dependent and energy-independent optical potentials making important comments in Sec. V. We apply in Sec. VI the new model to calculation
of the cross sections of the same $(d,p)$ reactions considered earlier  by us in Refs. \cite{Tim13b}. We discuss our model and the results obtained making conclusions in Sec. VII. Important derivations are given in the Appendix.
 
 \section{Connection between the three-body model of the $n+p+A$ system and the underlying many-body problem.}\label{Section II}
Explicit energy dependence of the nucleon optical potential arises as the result of internal structure of the target and the consequential possibility that the target can be excited. On the other hand, it is customary to model the $A(d,p)B$ reaction in terms of  scattering state solutions of a three-body Schr\"{o}dinger equation in which only the co-ordinates of the $n$ and $p$ appear explicitly.
We can formalize this 
procedure by regarding the three-body wavefunction of the model as the projection, $\Psi^{(+)}_{\ve{k}_d}(n,p)\phi_A$, of the full $A+2$ many-body wavefunction corresponding to a deuteron incident on a nucleus $A$ in its ground state, $\phi_A$, onto the ground state of  $A$. 
This 
projection 
has outgoing waves that describe elastic deuteron scattering and elastic deuteron break-up exactly and  outgoing waves in any open stripping channel in which $B$ has a non-neglible component with $A$ in its ground state. The complete many-body wavefunction has components in which $A$ is not in its ground state. The existence of coupling to these components influences the projection
 $\Psi^{(+)}_{\ve{k}_d}(n,p)\phi_A$ but we will base our analysis on a model in which their explicit contribution to the $A(d,p)B$ transition matrix is neglected.
 
 Within this framework an exact expression for $A(d,p)B$ transition matrix is 
 \begin{equation}\label{exact}
 T_{dp} = \langle \Psi_{\ve{k}_p,B}^{(-)}(p,B)|V_{np}|\Psi^{(+)}_{\ve{k}_d}(n,p)\phi_A\rangle,
\end{equation}
where $V_{np}$ is the $n-p$ interaction and $\Psi_{\ve{k}_p,B}^{(-)}$ is a solution of the many-body Schr\"{o}dinger equation corresponding to a proton with momentum $\ve{k}_p$ incident on $B$ but with $V_{np}$ deleted. This expression for the exact amplitude differs from the widely used alternative expression in which the transition operator is given by $V_{np}+\sum_{i\in A} V_{pi}-U_{pB}$ \cite{Satchler} and the bra-vector contains the product of the wave function of nucleus $B$ and the distorted wave in the $p-B$ channel generated by the (arbitrary) optical potential $U_{pB}$. The advantage of Eq. (\ref{exact}) is that it contains the short-ranged transition operator $V_{np}$ only, which  allows a simple approximation for the $\Psi^{(+)}_{\ve{k}_d}(n,p)$ to be developed. 
However, the final state wave function $\Psi_{\ve{k}_p,B}^{(-)}$ in this expression becomes more complicated. To further discuss this issue we restrict ourselves to a three-body model, $n+p+A$, in which the internal degrees of freedom of $A$ are not treated explicitly. In this case $\Psi_{\ve{k}_p,B}^{(-)}$  is a solution of a three-body problem, but it is a very special one in which (i) two of the bodies, $n$ and $p$, do not interact, and (ii) in practical applications to  $A(d,p)B $ reactions $A$ is frequently much more massive than a nucleon. In fact, in the $A\rightarrow \infty$ limit excitation of $B$ is impossible in this model and the exact solution for $\Psi_{\ve{k}_p,B}^{(-)}$  is the product of $\chi_{\ve{k}_p}^{(-)}(p)$, a proton scattering wave function distorted by the proton-target potential $V_{pA}$, and the wave function of the final nucleus $B$. The overlap of the latter with $A$ gives the neutron overlap function $I_{AB}$. Derivations of these results are given in \cite{RCJ09} where references to earlier work can be found. Corrections to this limit in the case of selected light nuclei $A$ have been evaluated in \cite{TJ99} using the adiabatic approximation to handle the excitation and break-up of $B$ by the recoil of $A$ and in \cite{Moro09} using the continuum-discretized coupled-channel method. It was shown in \cite{TJ99} how the recoil corrections modify the proton distorted wave, $\chi_{\ve{k}_p}^{(-)}(p)$, by a factor that goes to unity for large $A$ leaving a three-body wave function with a range limited in space by the neutron overlap function.  We refer to \cite{TJ99} and \cite{Moro09} for further quantitative discussion of these results which show that recoil contributions to  $\Psi_{\ve{k}_p,B}^{(-)}$ can probably be ignored in the first instance except for the very lightest nuclei. The aim of the present paper is to clarify how a proper treatment of the wave function in the incident channel changes $(d,p)$ predictions when the wave function in the final channel is fixed. Although to date a complete solution to the problem of calculating $\Psi_{\ve{k}_p,B}^{(-)}$ for finite $A$ has not yet been obtained the results of \cite{TJ99} and \cite{Moro09} give sufficient grounds for ignoring the finite $A$ complications in $\Psi_{\ve{k}_p,B}^{(-)}$ for our purposes.

The emphasis will now be on the evaluation of $\Psi^{(+)}_{\ve{k}_d}(n,p)$, which by definition is a function of the coordinates of $n$ and $p$. We will make heavy  use of the fact that the evaluation of the amplitude (\ref{exact}) requires knowledge of this function within the range of $V_{np}$ only and we will make approximations that are inspired by this observation.
    
Using projection operator techniques originally given by Feshbach \cite{Fes58} it is possible to derive  explicit expressions for the effective interactions that appear in the three-body  Hamiltonian that drives $\Psi^{(+)}_{\ve{k}_d}(n,p)$.  
 This effective 
Hamiltonian  is
  \begin{eqnarray}
 H_{\rm eff}=T_3+V_{np}+U^0_n(n)+U^0_p(p)+\langle \phi_A \mid U \mid \phi_A \rangle, \,\,\,\,\,\,\, \label{Heff}   
\end{eqnarray}
where $T_3$ is the three-body kinetic energy operator for $n+p+A$ in the centre of mass system and the bra-ket notation here implies integration over the target nucleus 
co-ordinates to leave an operator in $n$ and $p$ co-ordinates only. The interactions
$U^0_N(N)$ that appear in Eq. (\ref{Heff}) depend only on the $n-A$ and $p-A$ coordinates,
respectively, being diagonal in states of the target A, but otherwise arbitrary at this point. They
have no influence on the ground state matrix elements of $U+U^0_n+U^0_p$ or on the
three-body wave function.  The importance of the  introduction of the $U^0_N$'s in the present context is that their arbitrariness can be used to decouple the neutron and proton contributions to the complex many-body operator $U$, at least as far as the lowest multiple scattering contributions are concerned. How this is done will be discussed in Sec. IV.

The many-body operator $U$ in Eq. (\ref{Heff}) accounts for all the effects of the target nucleus degrees of freedom. It satisfies the integral equation (see Appendix A)
\begin{eqnarray}
U&=&(\Delta v_{nA}+\Delta_{pA}) 
+(\Delta v_{nA}+\Delta_{pA})\frac{Q_A}{e}U
\eol
\Delta  v_{NA} &=&  v_{NA} - U^0_N, \,\,\,\,\,\,
 v_{NA}=\sum_{i=1}^Av_{Ni}, \,\,\,\,\,\, N = n,p,\eol
\label{U}
\end{eqnarray}
The operator $Q_A$ projects onto excited states of the 
target and the energy denominator $e$ is given  by 
\begin{equation}e=E_3+i 0-T_3-V_{np}-U^0_n(n)-U^0_p(p) -(H_A-E_A),    \label{e} \end{equation}
 where $H_A$ and $E_A$ are the internal Hamiltonian  and the ground state energy of the target $A$, respectively. $E_3$ is the three-body energy related to the incident centre of mass kinetic energy $E_d$ and deuteron binding energy $\epsilon_0$ by $E_3=E_d-\epsilon_0$. 
 
 A formal solution of Eq. (\ref{U}) is
 \begin{eqnarray}
U&=& (\Delta v_{nA}+\Delta_{pA})
+(\Delta v_{nA}+\Delta_{pA}) Q_A\eol  &\times&
\frac{1}{e-Q_A(\Delta v_{nA}+\Delta_{pA})Q_A}(\Delta v_{nA}+\Delta_{pA}). \eol \label{U2}
\end{eqnarray}
 $\langle \phi_A \mid U \mid \phi_A \rangle$ sums up all processes via excited target states and the deuteron ground and break-up states that are coupled to the target ground state by  $\Delta v_{NA}$ and which begin and end 
on the target ground state. The expansion that appears on iteration of Eq. (\ref{U}) by repeated substitution for $U$ on the right-hand side makes this clear:
\begin{eqnarray}
U&=&(\Delta v_{nA}+\Delta_{pA}) \eol
&+&(\Delta v_{nA}+\Delta_{pA})\frac{Q_A}{e}(\Delta v_{nA}+\Delta_{pA}) 
\eol 
&+&(\Delta v_{nA}+\Delta_{pA})\frac{Q_A}{e}(\Delta v_{nA}+\Delta_{pA})\frac{Q_A}{e}(\Delta v_{nA}+\Delta_{pA}) 
\eol & +& \dots.  \label{Uiteration}
\end{eqnarray}
When averaged over the target ground state, and added to $U^0_n(n)+U^0_p(p)$, the first term on the right-hand-side of (\ref{Uiteration}) will be recognised as providing an expression for the sum of the real nucleon  optical potentials calculated in a folding model using the free-space interactions between $n$ and $p$ with the target nucleons. The second and higher order terms include the effect of target excitation. Because of the propagator $\frac{1}{e}$ they are complex and non-local in neutron and proton coordinates even when the nucleon-nucleon interactions $v_{Ni}$ are local. In addition there  is an explicit dependence on the energy $E_3$ through the energy denominator $e$. The purpose of this $E_3$ dependence is to link the three-body energy of the system to the thresholds of all relevant channels correctly. For example, if $E_3$ is less than the energy of a coupled intermediate excited state then the corresponding contribution to $U$ will be real (Hermitean); otherewise, a complex (non-Hermitean) contribution will result. This is the physical reason for expecting the effective interaction to be explicitly energy dependent.

Our aim is to make a connection between $U^0_n(n)+U^0_p(p)+\langle \phi_A \mid U \mid \phi_A \rangle$ and the neutron and proton optical potentials and, in particular, to determine the energies at which phenomenological optical models should be taken to reproduce the main properties of $U^0_n(n)+U^0_p(p)+\langle \phi_A \mid U \mid \phi_A \rangle$ for application to evaluation of the $(d,p)$ transition amplitude (\ref{exact}). We shall see that this can only be done approximately.

\section{Formal expressions for the nucleon optical potential in two- and three-body
channels.}

\subsection{Multiple scattering expansion of operator $U$.}\label{mult}

The operator $U$ which is  a complicated operator in the coordinates of $n$, $p$ and $ A$,  
can be partially separated into its $p-A$ and $n-A$ contributions by using manipulations 
from multiple scattering theory (see Appendix B). We obtain 
\begin{eqnarray}
U&=&U_{nA}+U_{pA} \eol
&+&U_{nA}\frac{Q_A}{e}U_{pA}+U_{pA}\frac{Q_A}{e}U_{nA}+.....,  \label{Umult}
\end{eqnarray}
where 
\begin{eqnarray}
U_{NA}= \Delta v_{NA} +\Delta  v_{NA}\frac{Q_A}{e}U_{NA}  \label{UNA}
\end{eqnarray}
and the dots in (\ref{Umult}) are terms of third or higher order in $U_{nA}$ 
and/or $U_{pA}$, always with an excited target (though not necessarily excited 
deuteron) as intermediate state. 

We emphasize that 
although the target ground state matrix element $U^0_n+U^0_p+\la \phi_A|U|\phi_A\ra$ is independent of the $U^0_N$ this is not true of the matrix elements $U^0_N+\la \phi_A|U_{NA}|\phi_A\ra$. The underlying reason for this is that the denominator $e$ in $U_{NA}$ depends both on $U_n^0$ and $U_p^0$ and, therefore, $U_{NA}$ is not symmetric in $n$ and $p$.
 
We  notice that the operators $U_{nA}$ and $U_{pA}$ in Eqs. (\ref{Umult}) and (\ref{UNA}) are strongly reminiscent of 
Feshbach's \cite{Fes58}  operator, which gives the nucleon 
optical 
potential when sandwiched between target ground state vectors. The exact expressions for this operator are given below.

\subsection{Two-body optical potential operator.}
  According to Feshbach \cite{Fes58} the  optical model operator for a nucleon with kinetic energy $E_N$ satisfies the integral equation
 \begin{eqnarray}
U^{\mathrm{opt}}_{NA}(E_N)=v_{NA} +v_{NA}\frac{Q_A}{e_N}U^{\mathrm{opt}}_{NA}(E_N),
 \label{UNAop}
\end{eqnarray}
 where $e_N=E_N+i 0-T_{NA}-(H_A-E_A)$. In terms of the solution of this equation the optical potential, $V^{\mathrm{opt}}_{NA}$, for nucleon scattering by $A$ at energy $E_N$ is given by
 \begin{eqnarray}
V^{\mathrm{opt}}_{NA}(E_N)= \langle \phi_A \mid U^{\mathrm{opt}}_{NA}(E_N)\mid \phi_A \rangle \eol
= \langle \phi_A \mid v_{NA}+v_{NA}Q_A\frac{1}{e_N-Q_Av_{NA}Q_A}v_{NA}\mid \phi_A \rangle.
 \eol \label{VnAop}
\end{eqnarray}
Note that in the proton case we include Coulomb interactions in the NN potenital $v_{pA}$ and, therefore, our proton optical potential includes a Coulomb potential $V_c(p) = 
\la \phi_A | \sum_{i=1}^Av^{coul}_{pi}| \phi_A \ra$.

To make a  connection with the $n+p+A$ case of the last subsection we introduce a slightly modified definition of nucleon optical potential:
\beq
\bar{U}^{\mathrm{opt}}_{NA} &=&
\Delta v_{NA}+\Delta v_{NA}\frac{Q_A}{\bar{e}_N}Q_A \bar{U}^{\mathrm{opt}}_{NA} 
\eol
&=&\Delta v_{NA}+\Delta v_{NA}Q_A\frac{1}{\bar{e}_N - Q_A \Delta v_{NA}Q_A}
\Delta v_{NA}, \eol
\eeqn{modU}
where 
\beq
\bar{e}_N = E_N + i0-T_{NA} - U^0_N -( H_A-E_A)
\eeqn{bareN}
differs from $e_N$ by inclusion of the potential $U^0_N$.  The operators 
$\bar{U}^{\mathrm{opt}}_{NA}$ and $U^{\mathrm{opt}}_{NA}$ have different matrix elements in general but their ground state expectation values are related  by
$\la \phi_A | \bar{U}^{\rm opt}_{NA}|\phi_A\ra = V^{\rm opt}_{nA} - U^0_N$.
The Feshbach optical potential
$V^{\mathrm{opt}}_{NA}$ is now given by
 \begin{eqnarray}
V^{\mathrm{opt}}_{NA}(E_N)=U^0_N(N)+\langle \phi_A \mid \bar{U}^{\mathrm{opt}}_{NA}(E_N)\mid \phi_A \rangle.
\label{VnAop}
\end{eqnarray}

Using the same basic relation between the nucleon optical potential and the underlying many-body theory that we use here Buck and Lipperheide \cite{Buc81,Buc83} showed that a non-local and energy \emph{independent} potential  operator can be formally defined that generates the same projection of the many-body wavefunction on to the target  ground state as does the explicitly energy dependent Eq. (\ref{VnAop}). However, the potential they derive as well as being non-Hermitean, as is also Eq. (\ref{VnAop}), does not satisfy  a standard  transformation under time reversal. When the underlying two-body interactions $v_{Ni}$ are invariant under time reversal,  Feshbach's   (\ref{VnAop}) satisfies
\beq \mathcal{K}V_{NA}^{\mathrm{opt}}\mathcal{K}^{-1}=( V_{NA}^{\mathrm{opt}})^{\dagger},\label{trev} \eeq
where $\mathcal{K}$ is the non-linear time reversal operator for the $N+A$ system. The operator defined by Buck and Lipperheide does not satisfy this relation. This results in a very clumsy discussion of reciprocity in reaction theories that make use of optical potentials and we therefore choose not to make use of the Buck-Lipperheide definition here.

\subsection {Relation between the three-body and two-body optical potentials.}

The three-body operator $U$  has contributions from the $U_{NA}$ as  leading term in the three-body Hamiltonian  
through Eqs.(\ref{Heff}) and (\ref{Umult}):
 \begin{eqnarray}
\langle \phi_A \mid U_{NA}(E_3) \mid \phi_A \rangle = \la \phi_A |\Delta v_{NA}|\phi_A\ra
\eol
+\langle \phi_A \mid \Delta v_{NA}Q_A\frac{1}{e-Q_A\Delta v_{NA}Q_A}
\Delta v_{NA}\mid \phi_A \rangle. \eol
\label{xxx}
\end{eqnarray}
These quantities have a similar structure to the Feshbach form for the nucleon optical potential discussed in the last Section. Note however that the 
energy denominator $e= E_3 +i0 -T_3 -V_{np}-U^0_n(n)-U^0_p(p)-(H_A-E_A)$ which appears everywhere in the $n+p+A$ case is not the denominator $\bar{e}_N= E_N +i0 -T_{NA}-U^0_N-(H_A-E_A)$ that appears in the neutron operator (\ref{bareN}). 

In order to deal with this difference 
when calculating $\la \phi_A | U | \phi_A\ra$ we proceed as follows:

\vspace{0.4 cm}
(i) We neglect the higher order multiple scattering terms that appear on the right-hand-side of Eq. (\ref{Umult}). These neglected terms involve  $n-A$ and $p-A$ excitations that cannot be expressed in any obvious way as a sum of terms depending separately on the $n$ and $p$ coordinates. They remain to be investigated in future work. These contributions are neglected in all three-body models of the $n+p+A$ system with heavy targets. They will not be discussed further.
  This means our approximation $U_{\rm eff}$ for the effective interaction  in the three-body Hamiltonian (\ref{Heff}) in addition to $V_{np}$ is
  \begin{eqnarray}
U_{\mathrm{eff}}(n,p)=
U_{\mathrm{eff}}^n(n)+U_{\mathrm{eff}}^p(p), \eol
U_{\mathrm{eff}}^N(N)=U^0_N(N)+\langle \phi_A \mid U_{NA}\mid\phi_A\rangle,
\label{Ueff}
\end{eqnarray}
where $N$ is $n$ or $p$ and the $U_{nA}$ and $U_{pA}$ are given by Eq. (\ref{UNA}).

\vspace{0.4 cm}
(ii) We consider only the
limit of the infinite target mass so that the three-body kinetic energy operator, $T_3$, can be separated into terms that depend on the individual neutron and proton coordinates relative to the target,  $T_3=T_p+T_n$, where $T_p \equiv T_{pA}$ and $T_n \equiv T_{nA}$. This approximation seems to be essential if we are to obtain a three-body Hamiltonian  which involves a sum of $n-A$ and $p-A$ potentials each of which only depend on the coordinates of one nucleon.  The energy denominator $e$ is now
\begin{eqnarray}
e=E_d+i0-\epsilon_0-T_n-T_p-V_{np}
\eol
-U^0_n(n)-U^0_p(p)-(H_A-E_A).  
\label{e2}
\end{eqnarray}

\section{Distorting potentials for the adiabatic model of $A(d,p)B$ reactions}\label{distortingpots}

 It has been shown \cite{PTJT} that for a range of incident deuteron energies of current experimental interest the $A(d,p)B$ amplitude (\ref{exact}) is dominated by the first Weinberg component of $\Psi^{(+)}_{\ve{k}_d}(n,p)$, \emph{i.e.},
 \begin{equation}\langle \phi_1 \mid \Psi^{(+)}_{\ve{k}_d}(n,p)\rangle, \label{chi1Psi}\end{equation}
 where $\phi_1$ is defined in terms of the $n$-$p$ interaction $V_{np}$ by
 \begin{equation}\mid \phi_1 \rangle=\frac{V_{np}\mid \phi_0 \rangle }{\langle \phi_0 \mid V_{np}\mid \phi_0 \rangle},\label{phi1}\end{equation}
in which $\phi_0$ is the deuteron ground state wavefunction. 
In (\ref{chi1Psi}), the integration is performed over the $n$-$p$ relative coordinate, $\ve{r}=\ve{r}_n-\ve{r}_p$, resulting in a function of the $n$-$p$ centre of mass coordinate, $\ve{R}=\frac{1}{2}(\ve{r}_n+\ve{r}_p)$, only. The state $\phi_1$ restricts the $n$-$p$ relative coordinate to be less than the range of the $n$-$p$ interaction $V_{np}$.

  In the ADWA approximation \cite{JT} the first Weinberg component of $\Psi^{(+)}_{\ve{k}_d}(n,p)$ is determined by a distorted wave $\chi^{(+)}_{\ve{k}_d}(\ve{R})$ satisfying the equation
 \begin{eqnarray}(E_d-T_R-\langle \phi_1 \mid U_{\mathrm{eff}}  \mid \phi_0\rangle)\chi^{(+)}_{\ve{k}_d}(\ve{R})=0,  \label{chiJT}\end{eqnarray}
  where $U_{\mathrm{eff}}$ is given by Eq. (\ref{Ueff}). 
We  first  consider the $U_{nA}$ term in the expression for $U_{\mathrm{eff}}$.
  
The dependence of the $U_{nA}$ on the proton coordinates is solely  through the presence in $e$ of the   operators $T_p+U^0_p$ and $V_{np}$. The values taken by $T_p+U^0_p+V_{np}$  influence   the net neutron energy parameter in $e$ and  hence determine  the appropriate energy at which to choose the neutron optical potential. Our method here is to replace $T_p+U^0_p+V_{np}$ by an average value consistent with dominance of the first Weinberg component to the $(d,p)$ reaction. We deduce this value within
the ADWA. Similarly, we will replace the $T_n+U^0_n+V_{np}$ operator in the $U_{pA}$ term for $U_{\rm eff}$ by an average value.

The ADWA potential in Eq. (\ref{chiJT}) has energy dependence that comes
through the energy denominators in $\la \phi_1 \phi_A | \Delta v_{NA} Q_A (e-Q_A\Delta v_{NA}Q_A)^{-1}Q_A\Delta v_{NA}|\phi_0 \phi_A \ra$ which contains
the matrix elements $\la  \phi_A | P_A \Delta v_{NA} Q_A| \phi_A\ra$ that are the functions
of $\ve{R}\pm\ve{r}/2$. The range of variable $\ve{r}$ is restricted
by the range of $\phi_1 \phi_0$ which is about 0.45 fm. Therefore, for most   $\ve{R}$
relevant for evaluating $\chi^{(+)}_{\ve{k}_d}(\ve{R})$ we have $r/2 \ll R$. Therefore, 
to a first approximation the operators $Q_A \Delta v_{NA}P_A$ in the numerator can be
replaced by their form at $\ve{r}=0$. Then the only matrix element involving integration over $\ve{r}$ is $\la  \phi_1 | (e-Q_A\Delta v_{NA}Q_A)^{-1}|\phi_0 \ra$.

  We show in Appendix C that within the ADWA  we can consistently use the approximation
\beq
\la  \phi_1 | (e-Q_A\Delta v_{NA}Q_A)^{-1}|\phi_0 \ra \approx
\frac{1}{\la  \phi_1 | e-Q_A\Delta v_{NA}Q_A|\phi_0 \ra}. \eol
\eeqn{oneovere}
We consider the contribution from $N=n$ first. To evaluate the denominator in the right-hand-side of (\ref{oneovere}) we use definition (\ref{e2}) for $e$ and the result from Appendix D
\beq
\la \phi_1|T_p+V_{np}|\phi_0\ra = \frac{1}{2}(T_R - \la T_r\ra) -\epsilon_0,
\eeqn{nav}
where $T_R$ and $T_r$ are the kinetic energy operators associated with the coordinates $\ve{R}$
and $\ve{r}$ respectively and
\beq
\la T_r\ra = \la \phi_1 | T_r |\phi_0\ra.
\eeqn{Tr}
We determine a reasonable value for the operator $T_R$ in Eq. (\ref{nav}) by recalling that it acts on the first Weinberg component $\chi^{(+)}_{\ve{k}_d}(\ve{R})$. Using  Eq. (\ref{chiJT}) we get
\beq 
\la  \phi_1 | e&-&Q_A\Delta v_{nA}Q_A|\phi_0 \ra \eol
&=& \frac{1}{2}[E_d + \la \phi_1 | U_{\rm eff}| \phi_0\ra  - 2 \la \phi_1 | U_p^0(p) | \phi_0 \ra
+\la T_r \ra] \eol
&-&[\la \phi_1 | T_n | \phi_0 \ra+ \la \phi_1 | U_n^0(n) | \phi_0\ra ] \eol
&-&\la \phi_1 | Q_A\Delta v_{nA} Q_A | \phi_0 \ra -(H_A-E_A).  
\eeqn{}
We now recall that the potentials $U_n^0$ and $U_p^0$ are arbitrary provided that they commute with $Q_A$. We are therefore at liberty to choose
\beq
\la \phi_1 | U_p^0(p) | \phi_0\ra  = \frac{1}{2}\la \phi_1| U_{\rm eff}|\phi_0\ra.
\eeqn{U0pav}
With this step Eq. (\ref{oneovere})  reduces to 
\beq
&\la & \phi_1 | (e-Q_A\Delta v_{nA}Q_A)^{-1}|\phi_0 \ra   \eol
=&\la & \phi_1| \frac{1}{{\cal E}_{\rm eff} - T_n - U_n^0(n) - Q_A\Delta v_{nA}Q_A}|\phi_0 \ra,
\eeqn{}
where
\beq
{\cal E}_{\rm eff} = \frac{1}{2}E_d + \frac{1}{2} \la T_r \ra.
\eeqn{eeff}
The neutron contribution to the ADWA effective interaction (\ref{Ueff}) is given by
\beq
& \la &\phi_1\mid  U_{\mathrm{eff}}^n(n)\mid \phi_0 \ra  \eol
&=&\la \phi_1 \mid U^0_n(n)+\la \phi_A \mid U_{nA}\mid\phi_A\ra \mid \phi_0 \ra \eol
&=&\la \phi_1 \mid U^0_n(n)\mid \phi_0 \ra 
 +\la \phi_1 \phi_A \mid \Delta v_{nA}+\Delta v_{nA}Q_A \eol
 &\times & \frac{1}{{\cal E}_{\mathrm{eff}}  - T_n 
-U^0_n(n)-Q_A\Delta v_{nA}Q_A}
\,Q_A\Delta v_{nA}\mid \phi_A \phi_0 \ra. \eol \label{UADWeffn}\eeq
We see that the expression (\ref{UADWeffn}) is just the neutron contribution to the ADWA distorting potential calculated with a non-local neutron optical potential taken at energy ${\cal E}_{\mathrm{eff}}$.
Similar conclusion can be made for proton contribution 
$ \la \phi_1\mid  U_{\mathrm{eff}}^p(n,p)\mid \phi_0 \ra$
to the ADWA distorting potential. With the choice 
\beq
\la\phi_1\mid U^0_n(n)\mid \phi_0\ra =\frac{1}{2}\la \phi_1\mid U_{\rm eff}\mid \phi_0\ra
\eeqn{U0nav} 
we find in identical fashion that  this contribution  is obtained using the $p-A$ optical potential, including Coulomb terms, taken at the proton energy $E_p$ also
equal to ${\cal E}_{\rm eff} = \frac{1}{2}(E_d+\la T_r\ra )$. We note that with the choices
given by Eqs. (\ref{U0nav}) and (\ref{U0pav})  the neutron and proton contributions to the effective interaction of the three-body model in the single scattering approximation are independent of the $U_N^0$.  The choice (\ref{U0nav}) for $U_n^0$ will influence higher order multiple scattering terms in Eq. (\ref{Umult}). We will not consider these terms further here.

\section{Comments}

We have given arguments that suggest that for the purpose of calculating $(d,p)$ cross
sections in the ADWA approximation  when the  nucleon optical potentials are explicitly energy dependent as well as non-local  the neutron and proton kinetic energies used in the incident channel should be $\frac{1}{2}E_d +\frac{1}{2}\la T_r\ra$. This is acheved by averaging
the proton kinetic energy in the incoming deuteron and adding the extra kinetic energy the neutron has because it must be close to proton in order to contribute to the $(d,p)$
amplitude. Thus the problem of calculating the ADWA amplitude for $A(d,p)B$  with energy-dependent nonlocal potentials is reduced to the problem of calculating this amplitude with
energy-independent nonlocal potentials, the solution of which is found in Refs. \cite{Tim13a,Tim13b}. If a phenomenological non-local, explicitly energy dependent nucleon potential is available, the  prescription of present paper together with that
from Refs. \cite{Tim13a,Tim13b} can be used unambiguously.  

The energy-independent non-local potential that replaces the original energy-dependent nonlocal potential is obtained from the latter by evaluating  it at the energy shifted from
the intuitively assumed $E_d/2$ value by one half of the average $n-p$ kinetic energy within
the range of the $n-p$ potential, approximately equal to 57 MeV \cite{Tim13a}. 
This value is close to, but not equal to,  the energy shift   $\Delta E \sim 40 $ MeV, identified in Refs. \cite{Tim13a,Tim13b} as the shift to the $E_d/2$ value
that provides the appropriate energy of energy-dependent equivalents of nonlocal potentials
to be used in ADWA.

For  all known  phenomenological optical  potentials the underlying
energy-dependent non-local potentials are not known. It may be tempting to use
the conclusion of the previous section for choosing in the ADWA calculations the
phenomenological local potentials taken at energies   ${\cal E}_{\rm eff} = \frac{1}{2}E_d +\frac{1}{2}\la T_r\ra$.
This, however, would be a source of confusion and may lead to a wrong result.
Indeed, if ${\cal U} (E,r,r')$ is the underlying energy-dependent non-local potential for the
phenomenological local energy-dependent equivalent $U_{phen}(E,r)$ then the ADWA needs to use
the energy-independent non-local potentials ${\cal U}({\cal E}_{\rm eff},r,r')$. This would give a
different equivalent energy-dependent  potential $ {\tilde U} (E,r)$ which then should be
taken at the energy $E_d/2+\Delta E$, as shown in \cite{Tim13a}, \cite{Tim13b}. The two potentials $ {\tilde U} (E_d/2+\Delta E,r)$
and $U_{phen}({\cal E}_{\rm eff},r)$ are not the same. We will derive a link between them for a particular class of ${\cal U} (E,r,r')$ in the next section. However, more generally, further knowledge of energy-dependent non-local potentials is needed.

It is interesting to compare our treatment of explicit energy dependent optical potentials with that of Ref. \cite{Del09a}.
In that work the Faddeev approach was used to solve the three-body problem.  Energy dependence is treated  in \cite{Del09a}   by taking the neutron (proton)  optical
potential  at an energy equal to the variable neutron (proton) energy parameter used   in 
evaluating the neutron (proton) $t$-matrix as it appears in the Faddeev equations. This seems physically plausible at first sight, but what kind of Schr\"odinger equation such an approach corresponds to is not clear. The Faddeev equations have been derived from the Schr\"odinger equation with energy-independent non-local pairwise potentials. We are not aware of a formal derivation of the Faddeev equations for explicitly energy-dependent pair potentials of the type that arise from many-body effects as discussed in Section \ref{Section II}.

In our view, explicit energy dependence
is an effect that can only be understood by explicitly recognising the internal degrees of freedom of the target as is done here and in Refs. \cite{Muk12,Joh05}. It should be added that the approximate prescription proposed here is intended for use only in the calculation of the incident channel distorting potential in a particular expression for the $A(d,p)B$ transition matrix and is not expected to be useful in any other context.

\section{Applications to the $^{16}{\rm O}(d,p)^{17}{\rm O}$, 
$^{36}{\rm Ar}(d,p)^{37}{\rm Ar}$ and $^{40}{\rm Ca}(d,p)^{41}{\rm Ca}$ reactions
with global energy-dependent nonlocal optical potential}

In this section we apply the formalism developed above to the ADWA calculation of the
cross sections of the 
$^{16}{\rm O}(d,p)^{17}$O, $^{36}{\rm Ar}(d,p)^{37}$Ar and $^{40}{\rm Ca}(d,p)^{41}$Ca
reactions at $E_d = 15$, 9 and 11.8 MeV respectively. The cross sections of these
reactions have been measured in \cite{o16,ar36,ca40}. These are the same
 reactions that have been considered   in Ref. \cite{Tim13b} using the
energy-independent non-local nucleon optical potentials given by Giannini and Ricco (GR)
systematics in   \cite{Gia76}.
The choice of targets made in \cite{Tim13b} was to justify the use of the GR potential which is  valid for $N=Z$ nuclei only.
In the present work, we first use the  Giannini-Ricco-Zucchiatti (GRZ) nonlocal potential \cite{GRZ}
which has an energy-dependent imaginary part. Then we make a suggestion of how to account for non-locality when only energy-dependent local phenomenological potentials are known but the underlying energy-dependent non-local potentials are not known. As an example we use the widely used CH89 systematics \cite{CH89}.

\subsection{Giannini-Ricco-Zucchiatti potential}

For $N=Z$
targets, the GRZ potential is given by
\beq
{\cal U}_{NA}(E,\ve{r},\ve{r}')=H(\ve{r}-\ve{r}')U_{NA}(E,(\ve{r}+\ve{r}')/2), 
\eeqn{GRZ}
with the nonlocality factor  
\beq
H(\ve{x}) = \pi^{-3/2} \beta^{-3} e^{-\left(\frac{\ve{x}}{\beta}\right)^2}
\eeqn{HPB}
determined by the non-locality range $\beta = 2\hbar^2\alpha/\mu$, where $\alpha = 0.0116$ MeV$^{-1}$ and $\mu$ is the nucleon-target reduced mass. The potential formfactor 
$U_{NA}$ is defined as
\beq
U_{NA}(E,y) &=& -V_N f_N(y)-4iW_N(E)f_{NI}(y)(1-f_{NI}(y)), 
\eol
f_i(y)&=&(1+\exp((y-R_N)/a_i))^{-1}.
\eeqn{}
It has one energy-dependent parameter, the depth of the imaginary part,
\beq
W_N(E) = 17.5 (1-\exp(-0.05E)) \,\, {\rm MeV}.
\eeqn{W}
All other parameters are  energy-independent and have the following values: the depth of the real potential $V_N = 85 $ MeV, the radius of both the real and imaginary potentials $R_N = 1.16A^{1/3}$ fm, the diffusseness $a_N = 0.57$ fm and $a_{NI} = 0.54 + 0.0032 A$ fm 
for the real and the imaginary potentials respectively. We neglect the spin-orbit term because its contributions to the $(d,p)$ cross sections at the chosen incident deuteron energies  are small.

\begin{figure*}[t]
\vspace{0.4 cm}
\centering
{\includegraphics[width=0.31\textwidth]{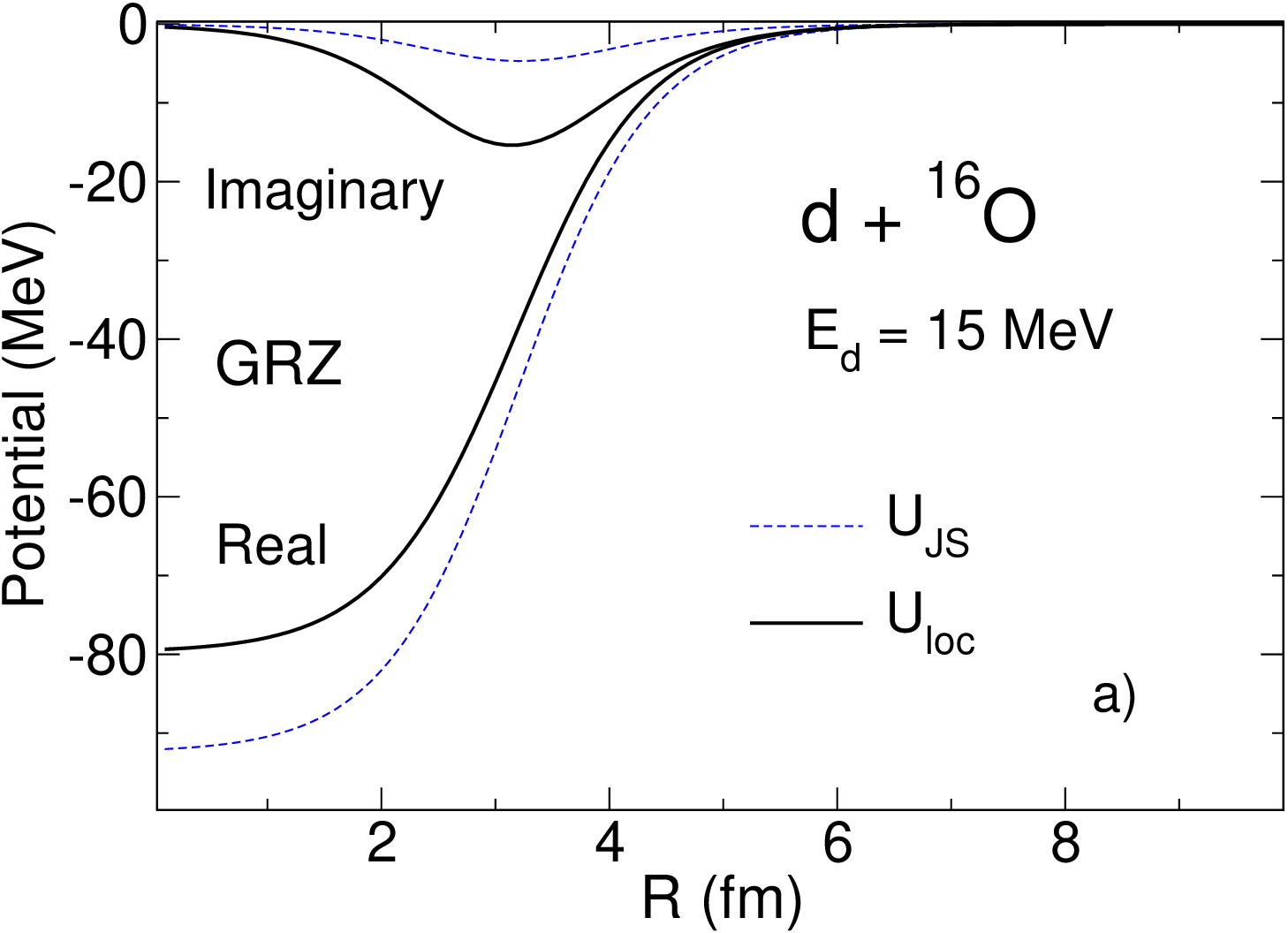} \,\,\,\vspace{0.3 cm}
\includegraphics[width=0.31\textwidth]{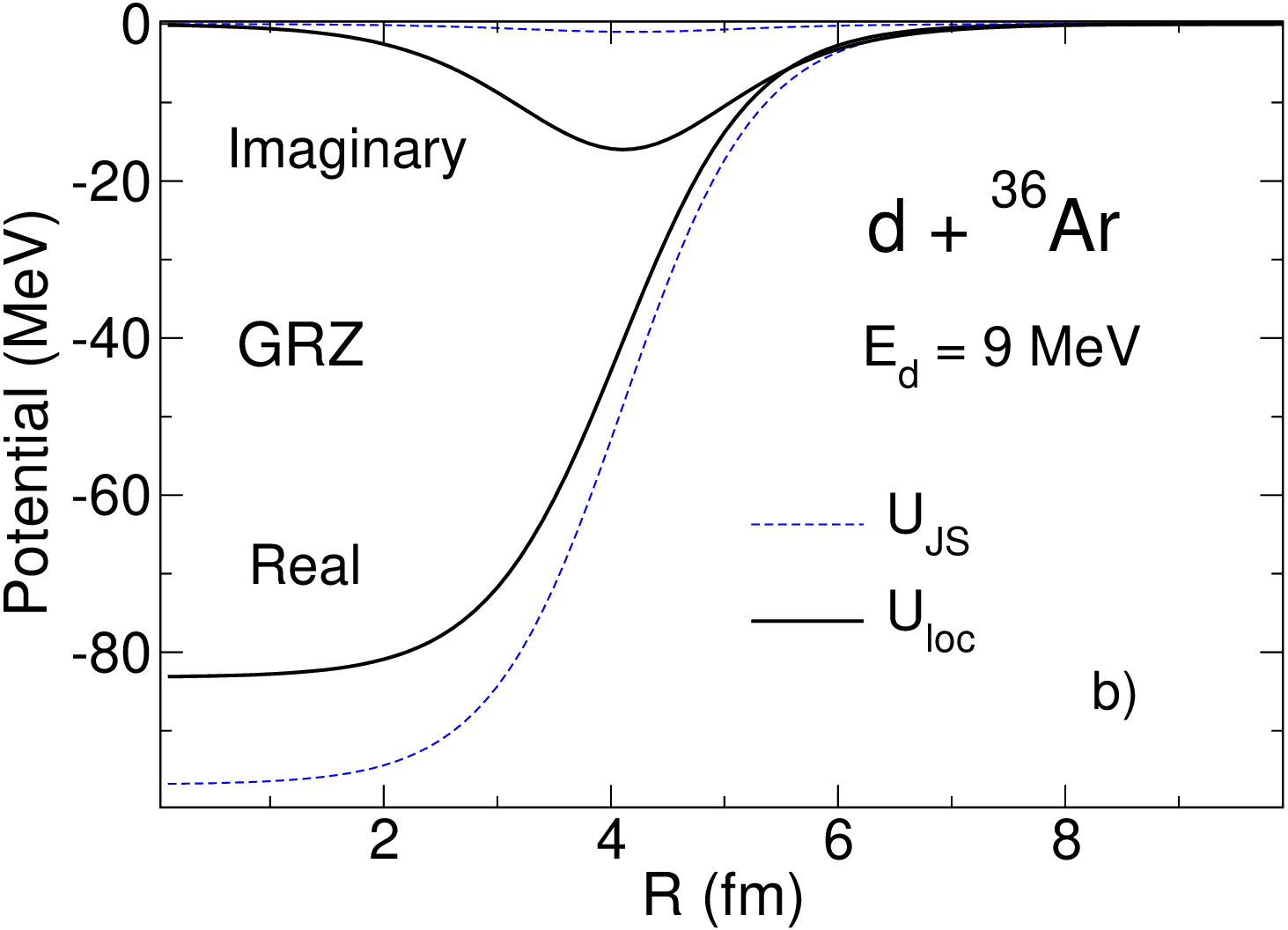} \,\,\,\vspace{0.3 cm}
\includegraphics[width=0.31\textwidth]{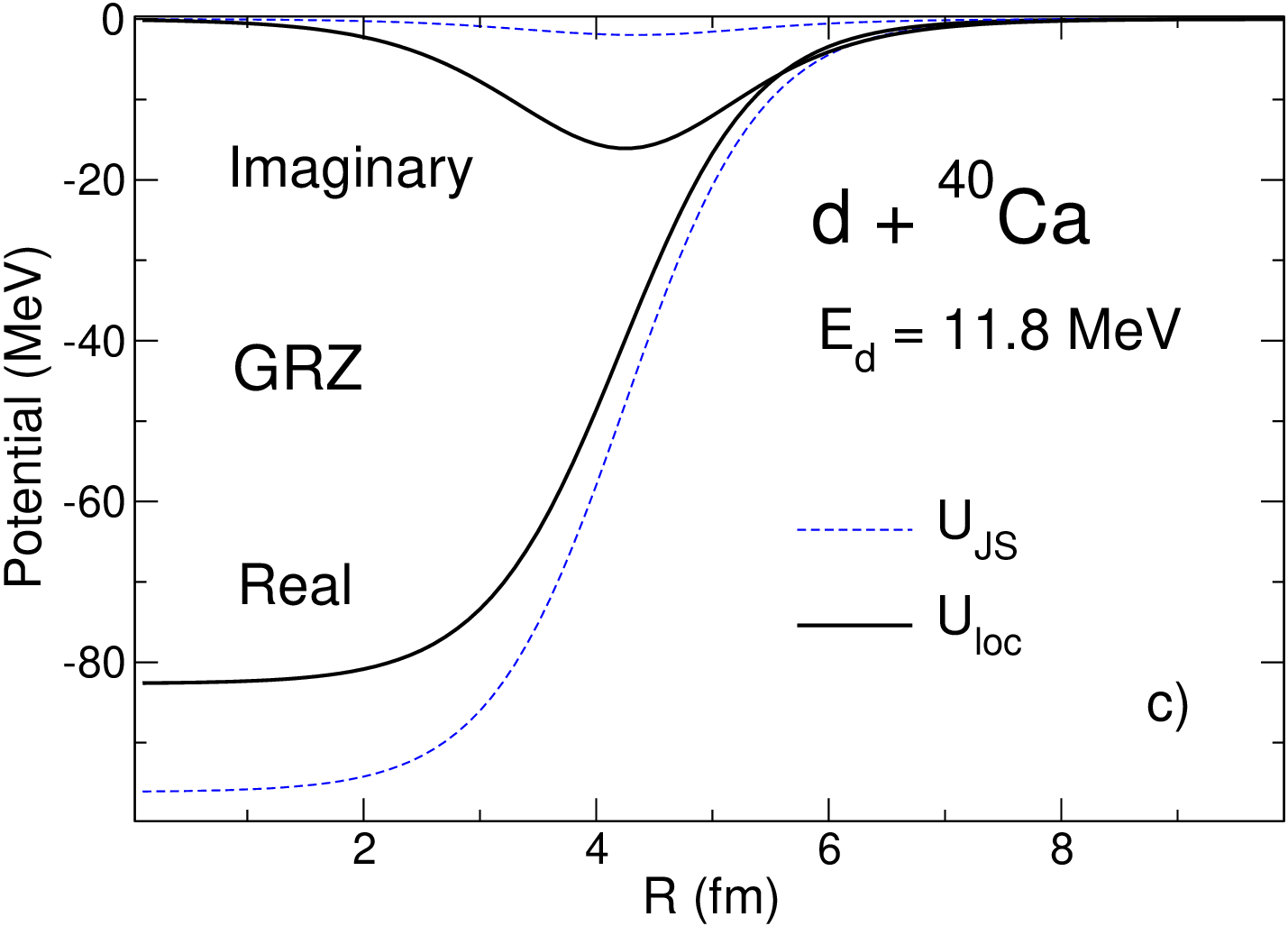}} \vspace{0.3 cm}
\includegraphics[width=0.31\textwidth]{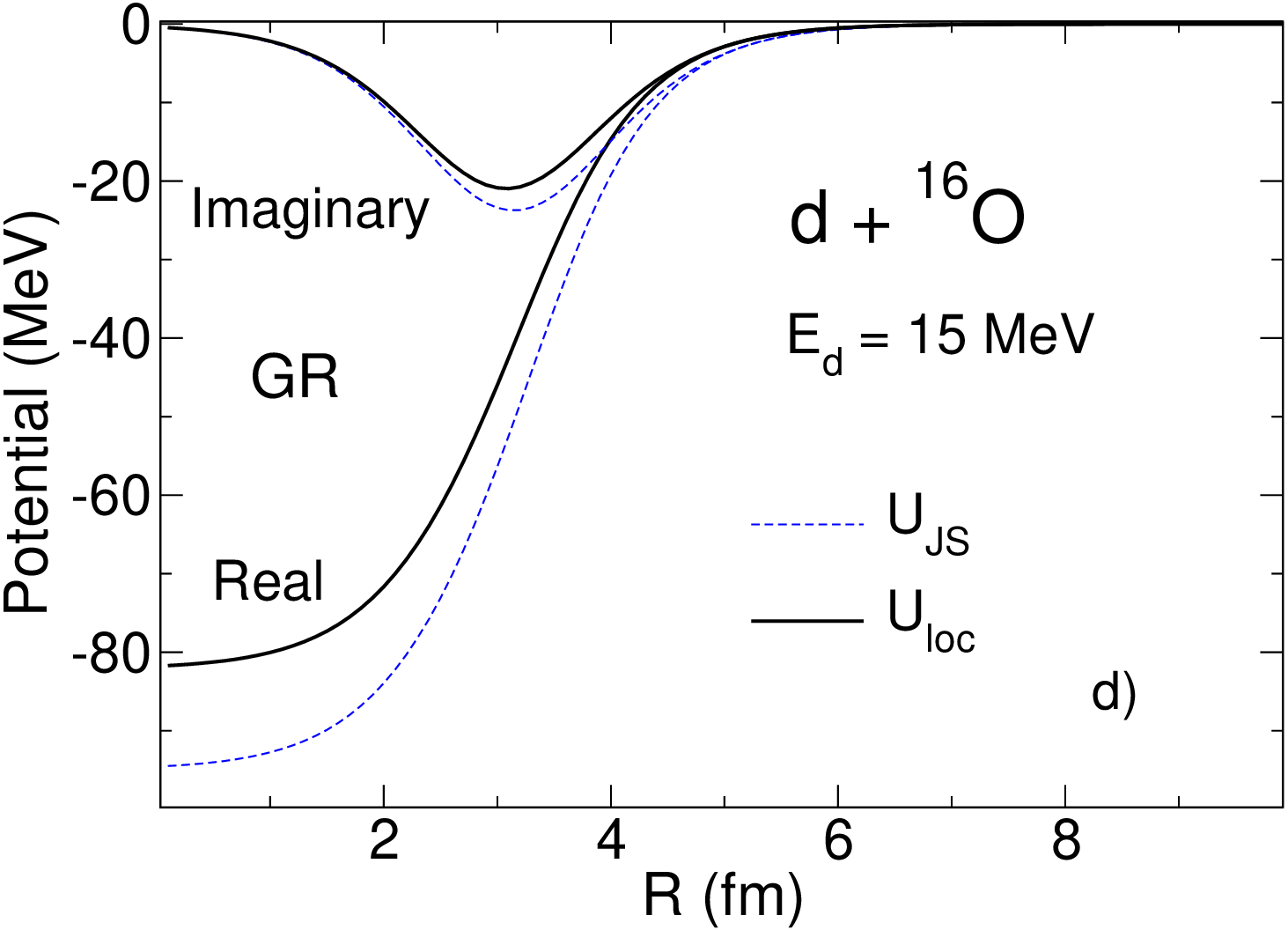} \,\,\,\vspace{0.1 cm}
\includegraphics[width=0.31\textwidth]{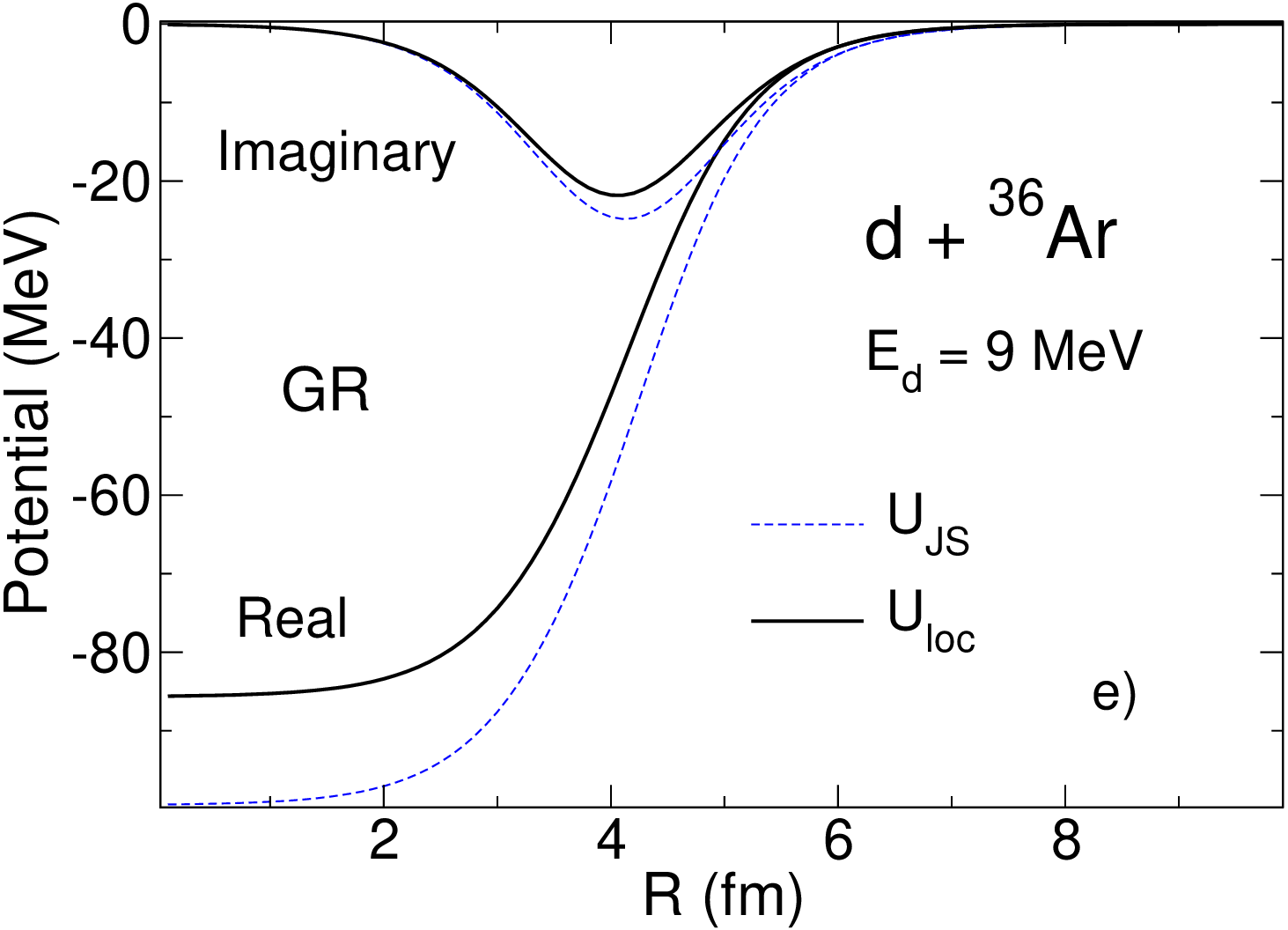} \,\,\,\vspace{0.1 cm}
\includegraphics[width=0.31\textwidth]{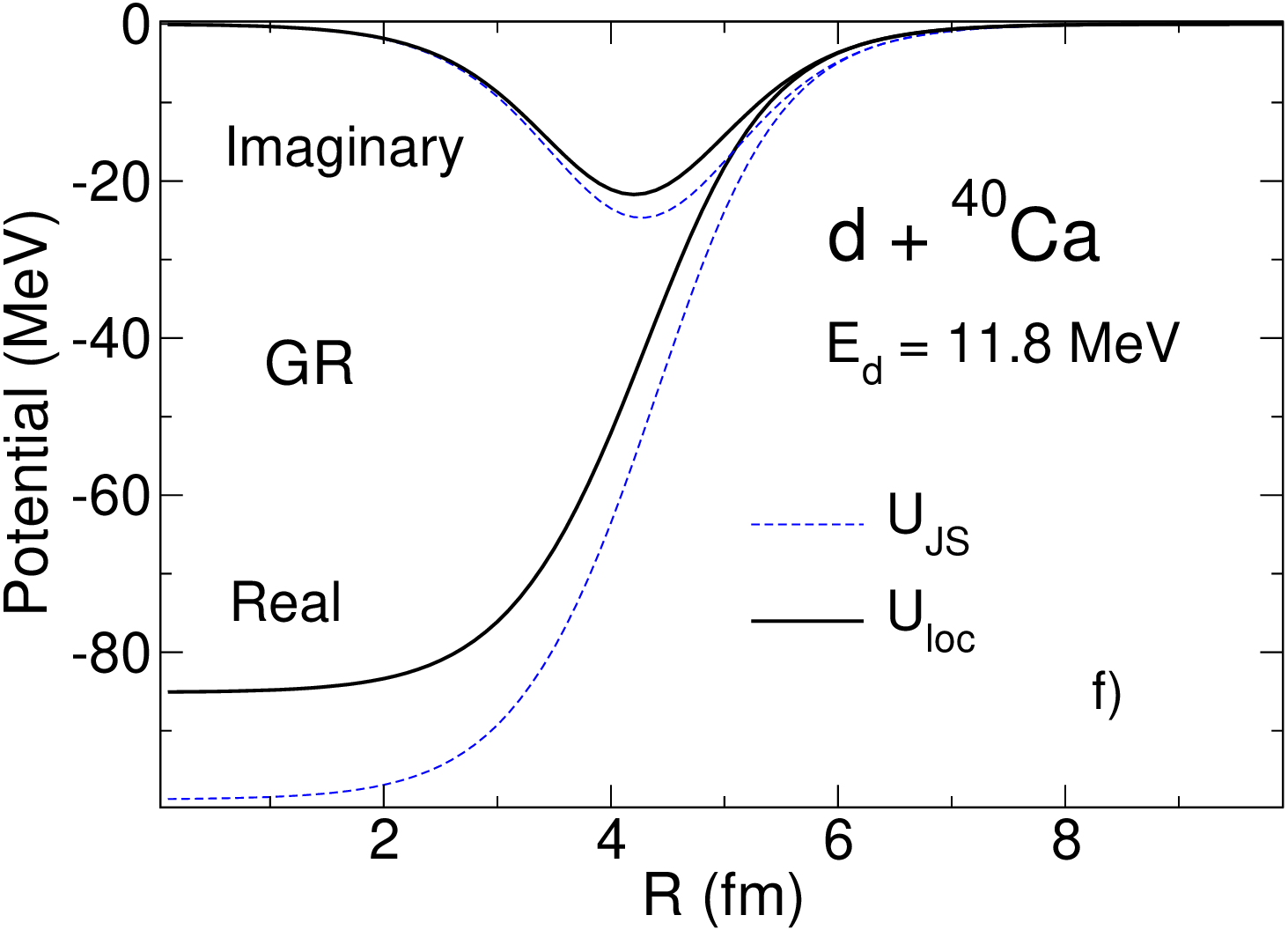}\vspace{0.1 cm}
\includegraphics[width=0.31\textwidth]{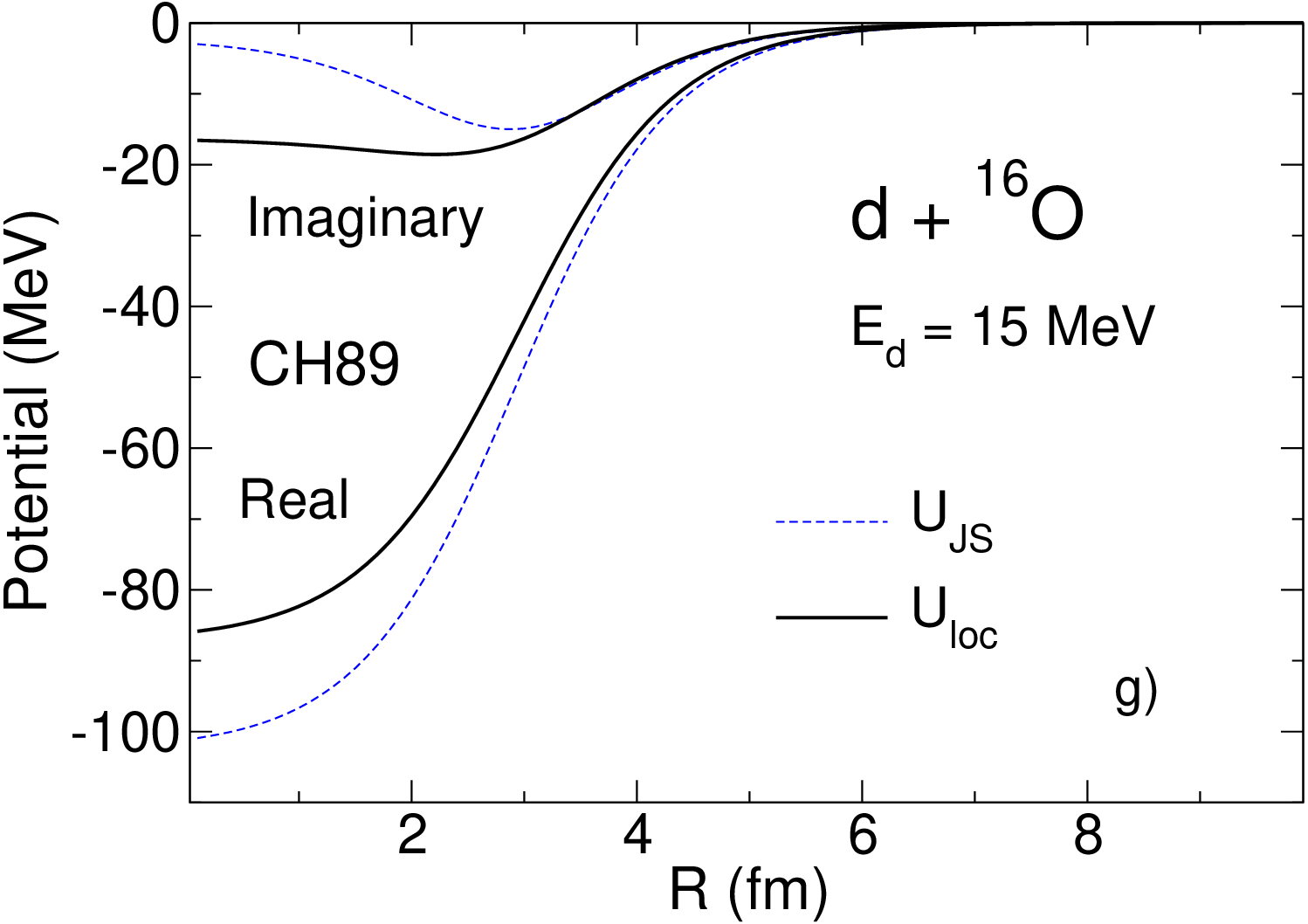} \,\,\,
\includegraphics[width=0.31\textwidth]{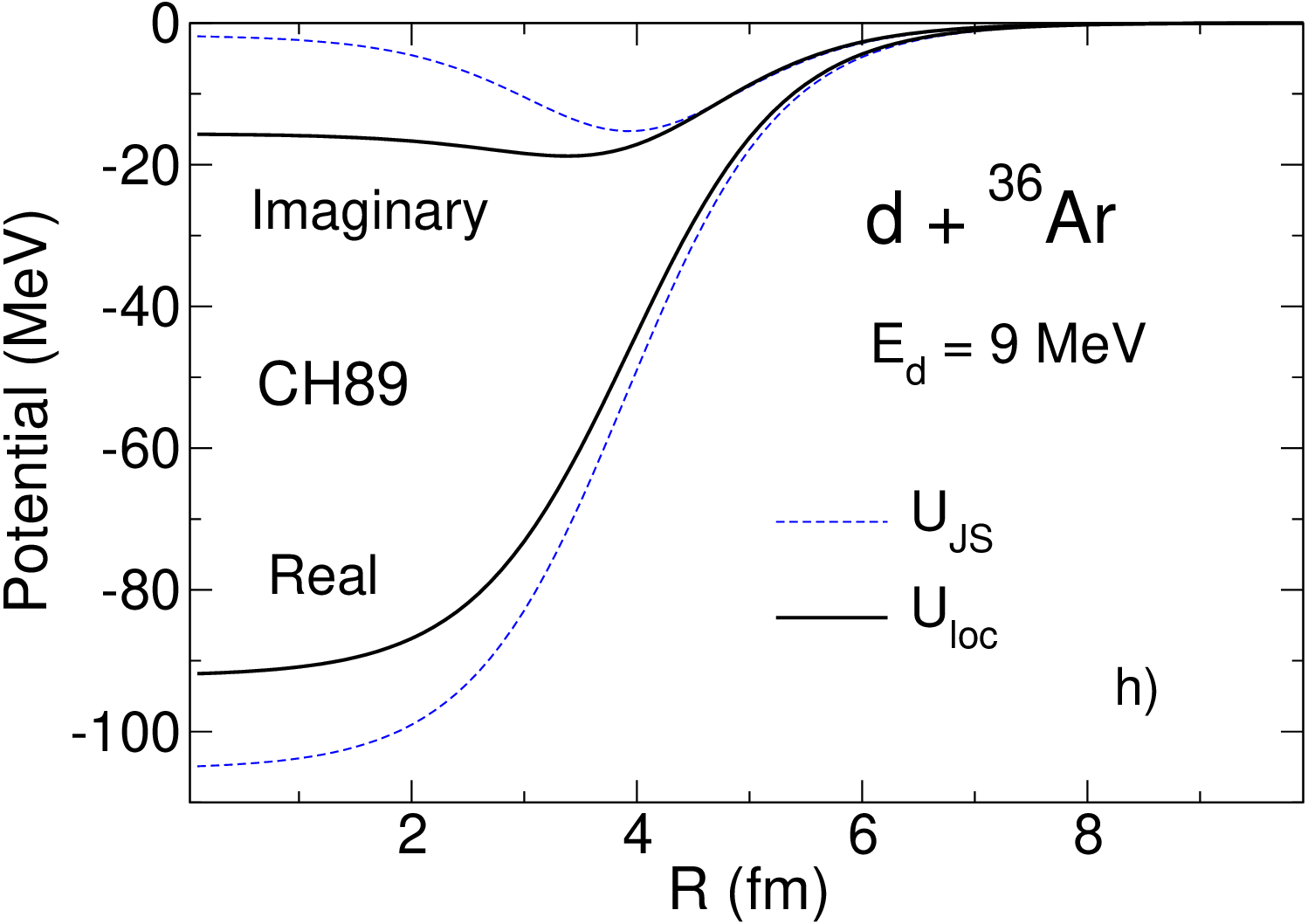} \,\,\,
\includegraphics[width=0.31\textwidth]{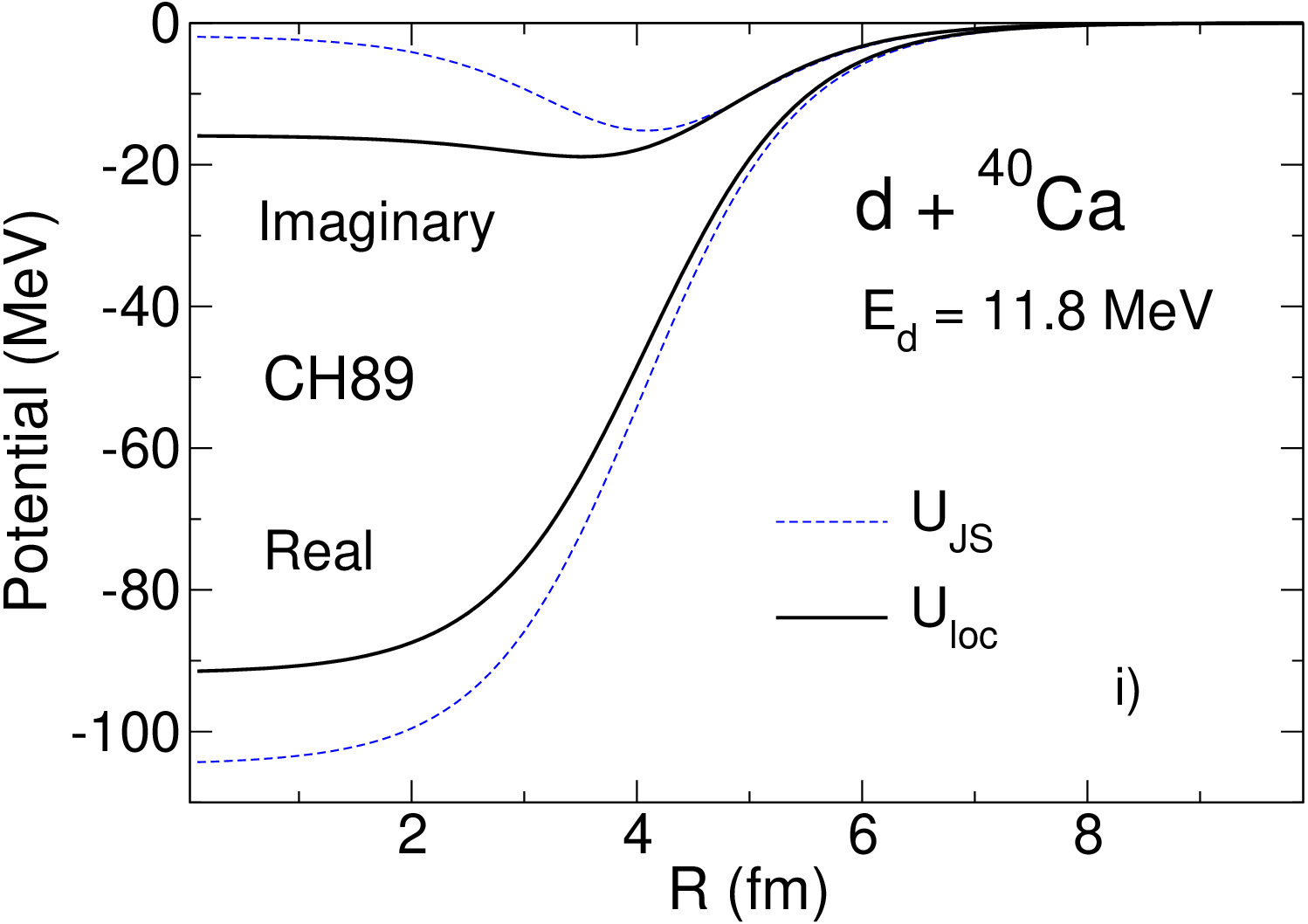}
 \caption{(Color online) 
 Local adiabatic deuteron  potentials $U_{loc}$ 
 for  energy-dependent Giannini-Ricco-Zucchiatti (GRZ) ($a,b,c$),  energy-independent Giannini-Ricco (GR) 
 ($d,e,f$) non-local optical nucleon potentials and energy-dependent local potential
 CH89 ($g,h,i$) in comparison  
 to the corresponding Johnson-Soper potentials $U_{JS}$ for $^{16}$O, $^{36}$Ar
 and $^{40}$Ca targets. 
 }\label{fig2}
\end{figure*}
\begin{figure*}[t]
\vspace{0.4 cm}
\centering
{\includegraphics[width=0.32\textwidth]{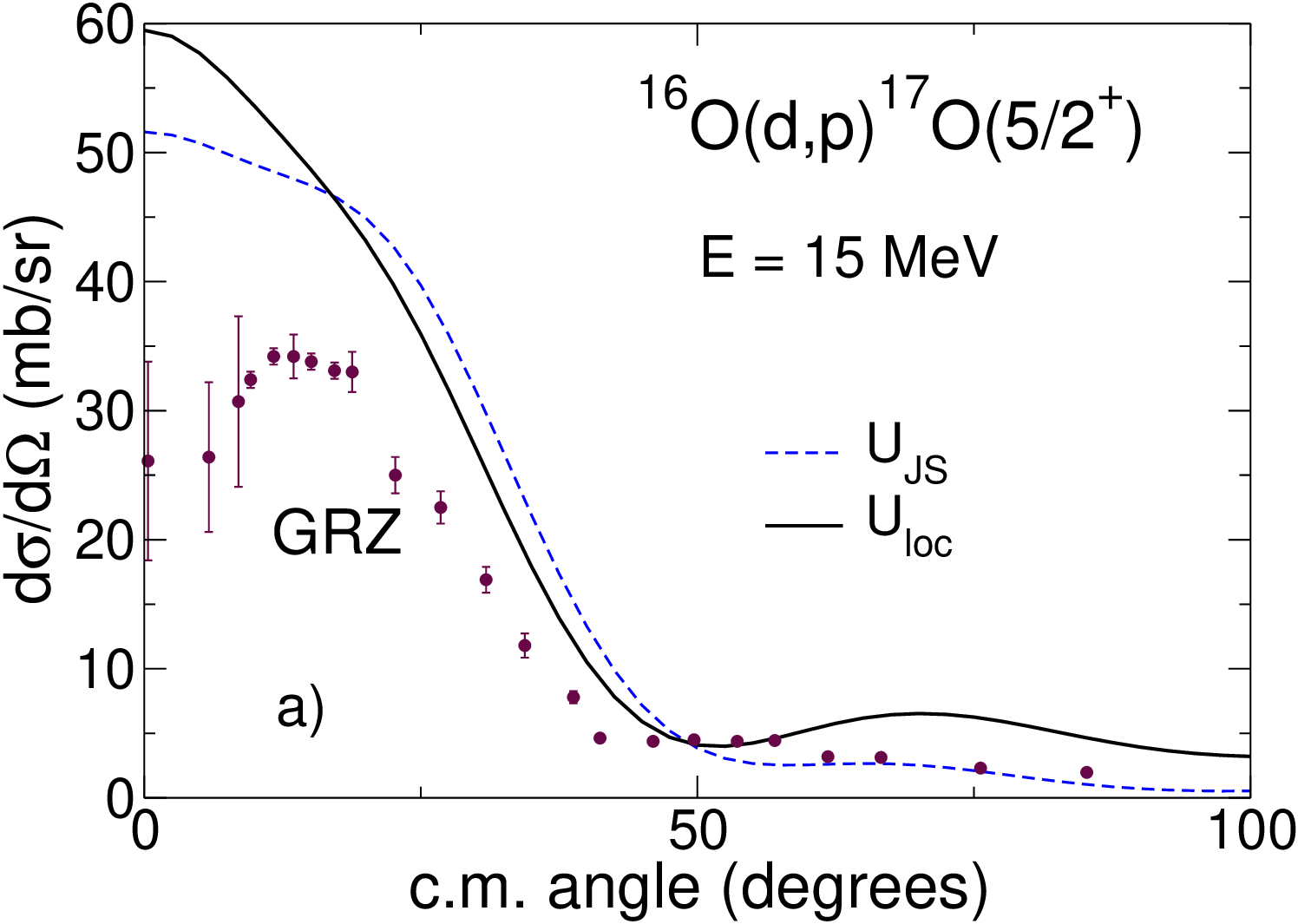} \,\,\,\vspace{0.3 cm}
\includegraphics[width=0.315\textwidth]{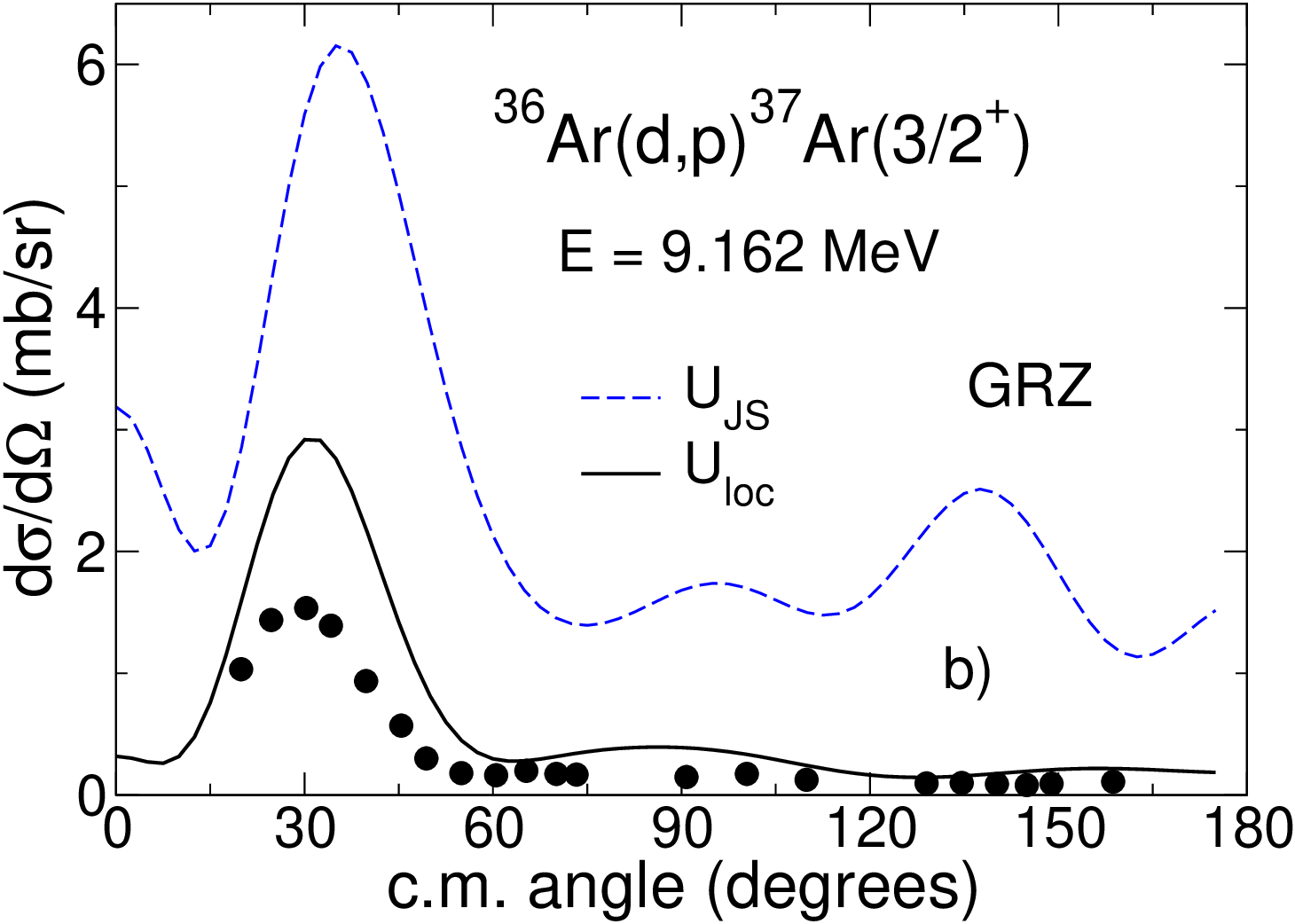} \,\,\,\vspace{0.3 cm}
\includegraphics[width=0.32\textwidth]{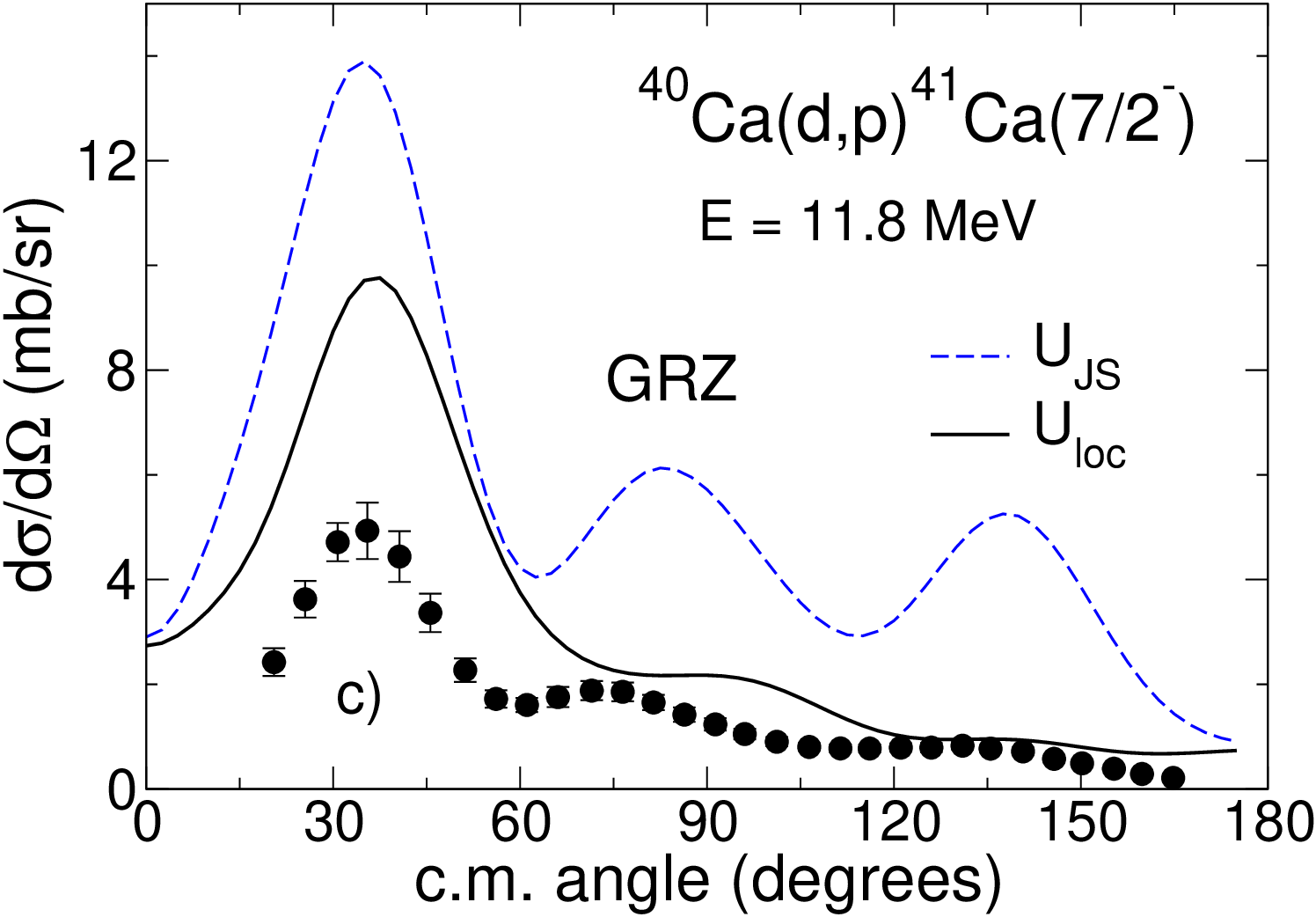}} \vspace{0.3 cm}
\includegraphics[width=0.32\textwidth]{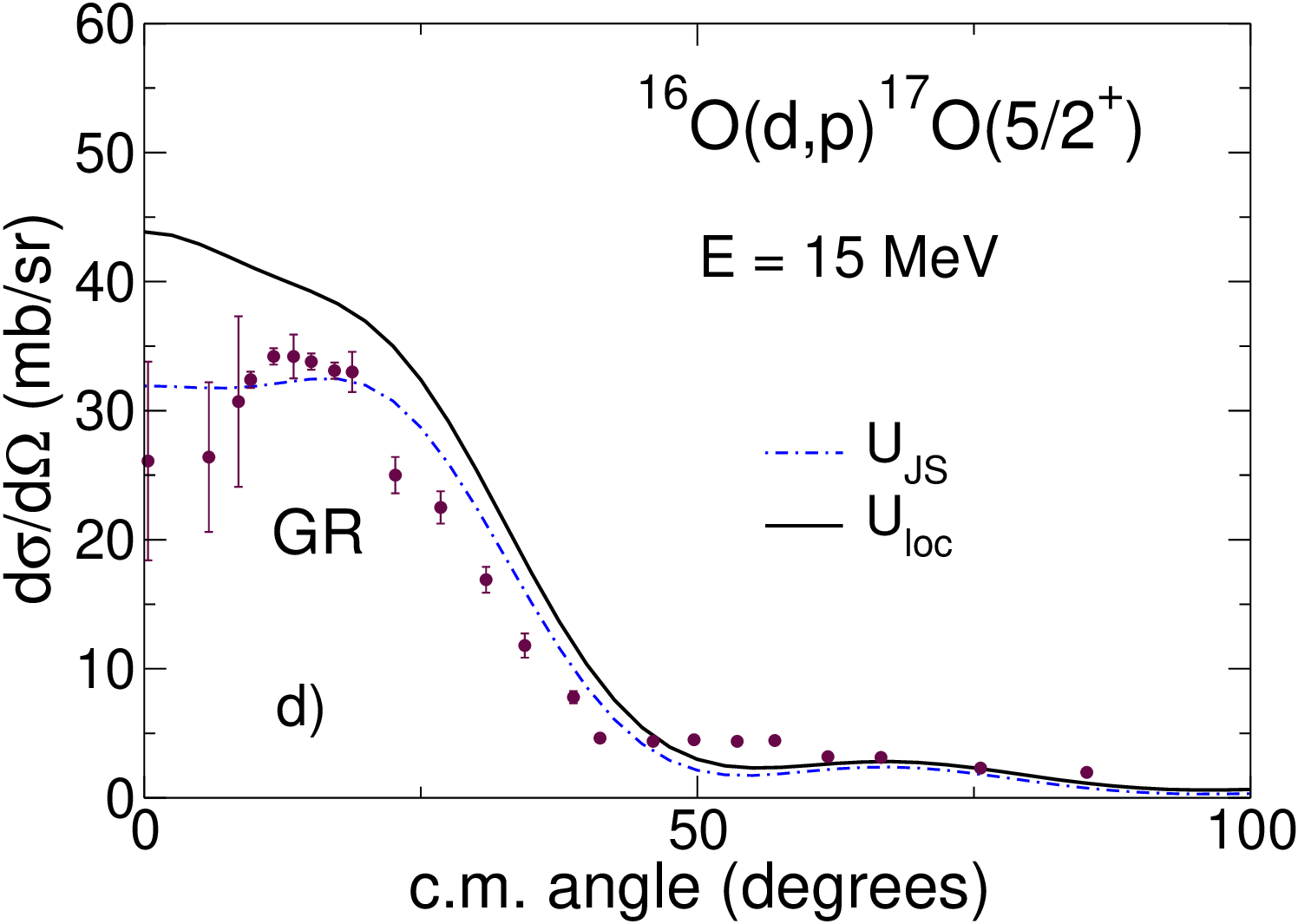} \,\,\,\vspace{0.1 cm}
\includegraphics[width=0.315\textwidth]{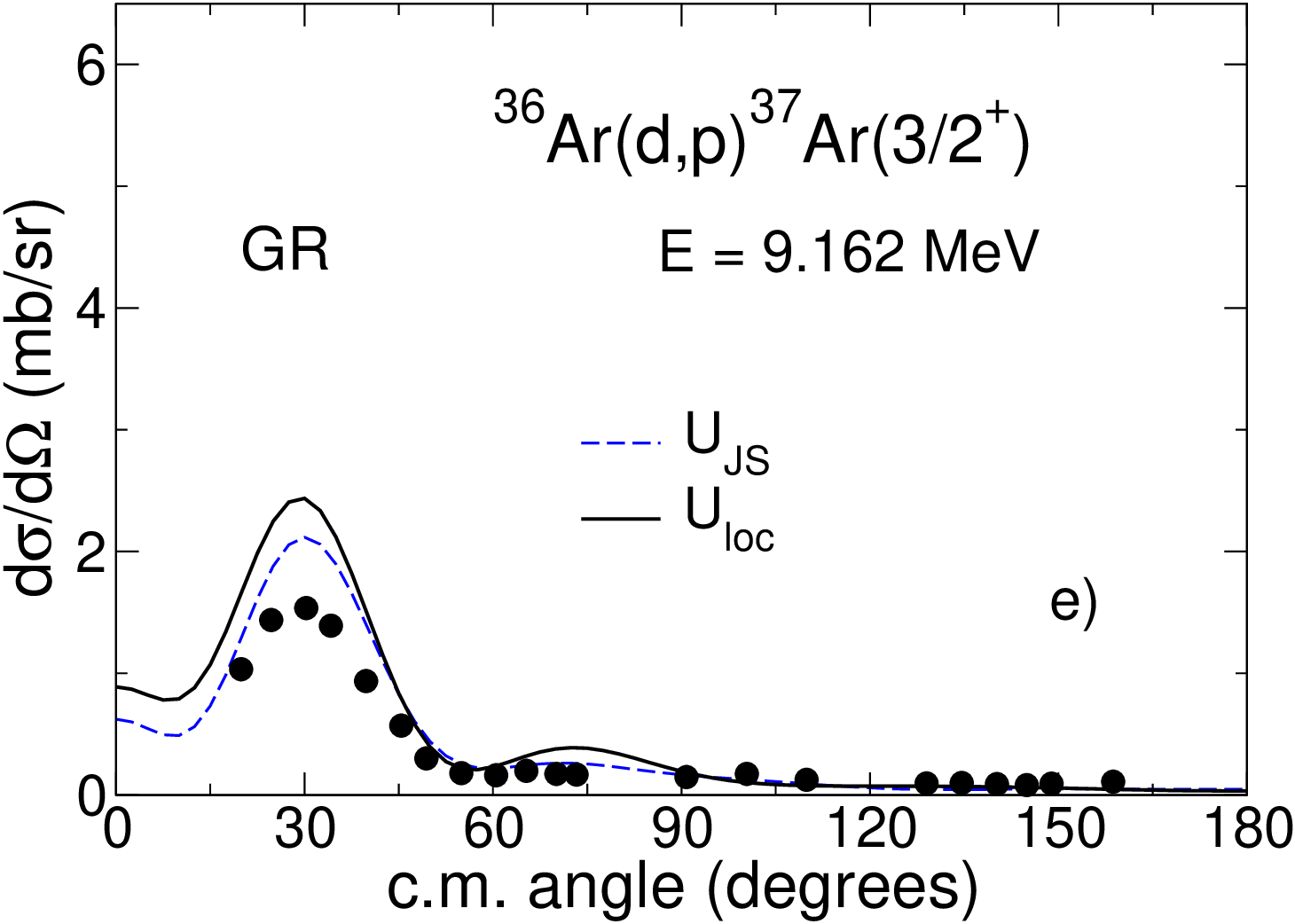} \,\,\,\vspace{0.1 cm}
\includegraphics[width=0.32\textwidth]{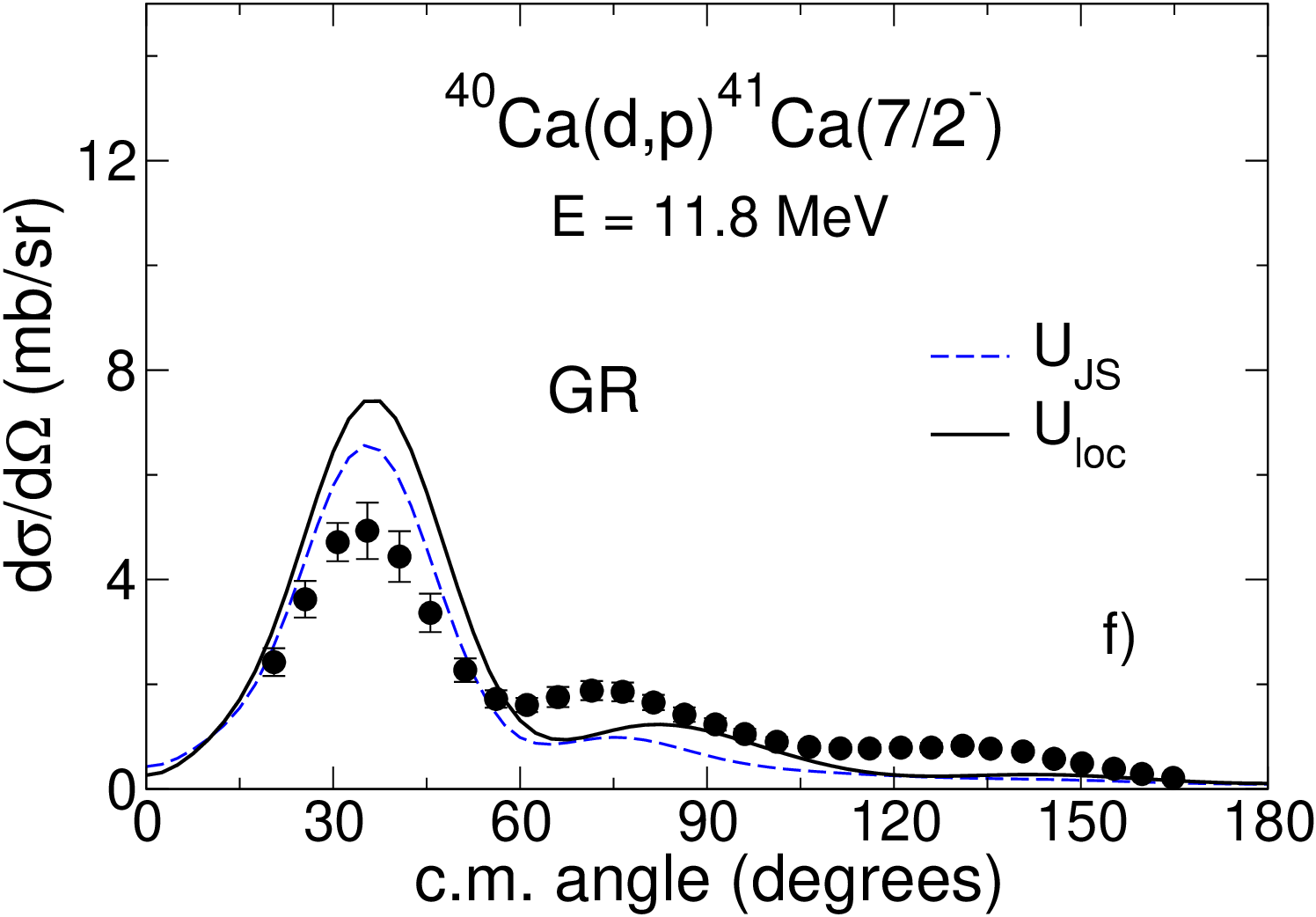}\vspace{0.1 cm}
\includegraphics[width=0.32\textwidth]{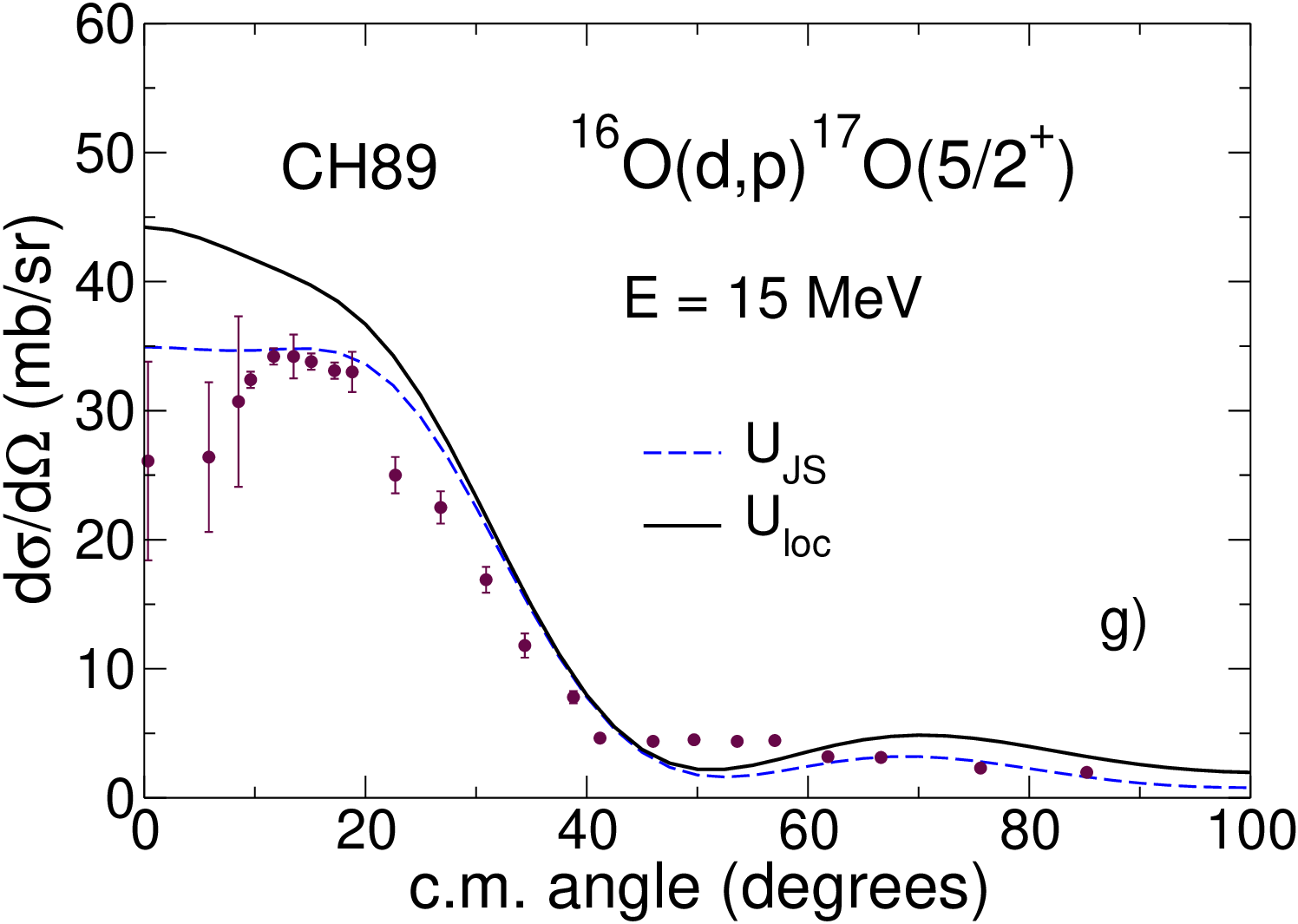} \,\,\,
\includegraphics[width=0.315\textwidth]{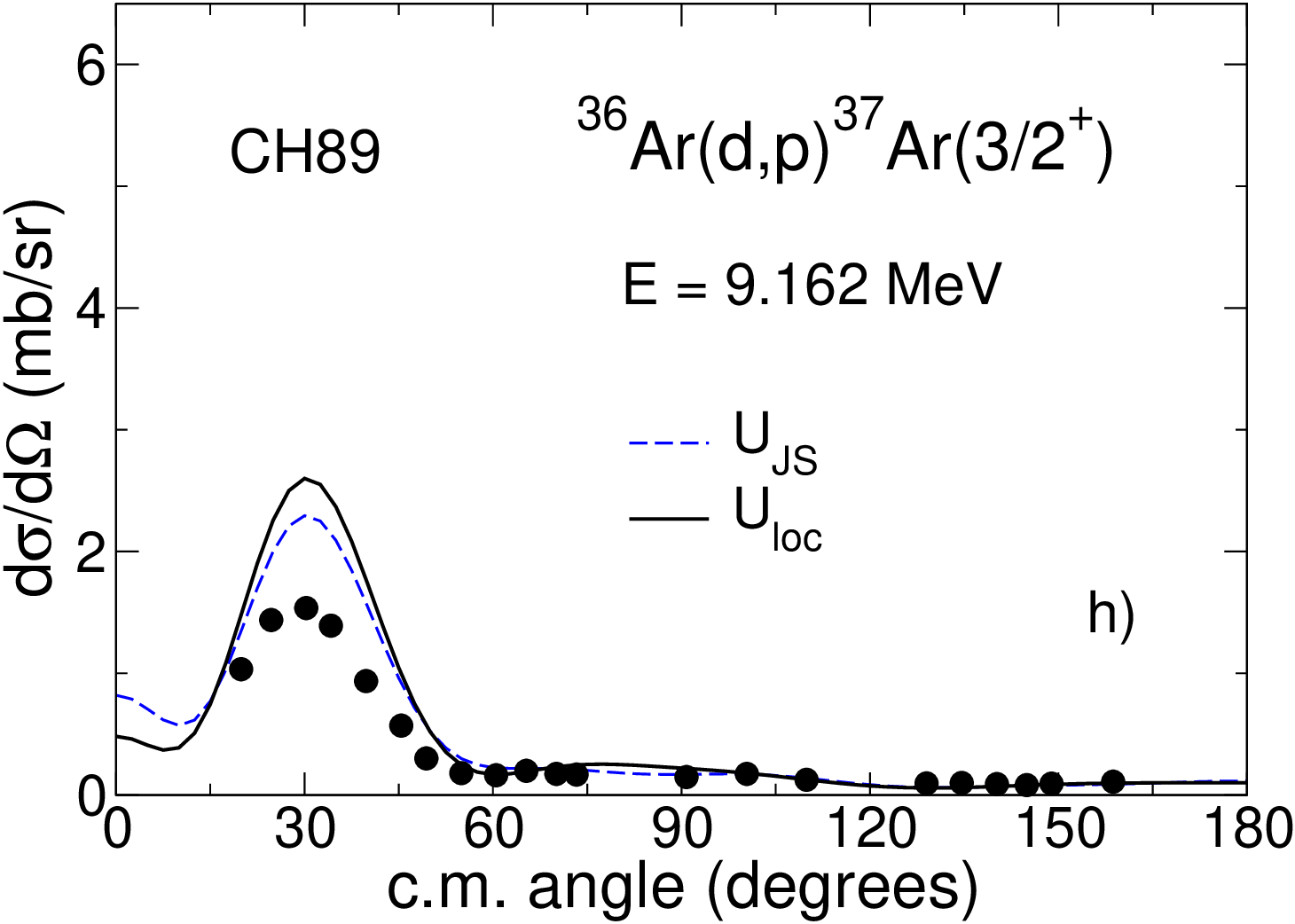} \,\,\,
\includegraphics[width=0.32\textwidth]{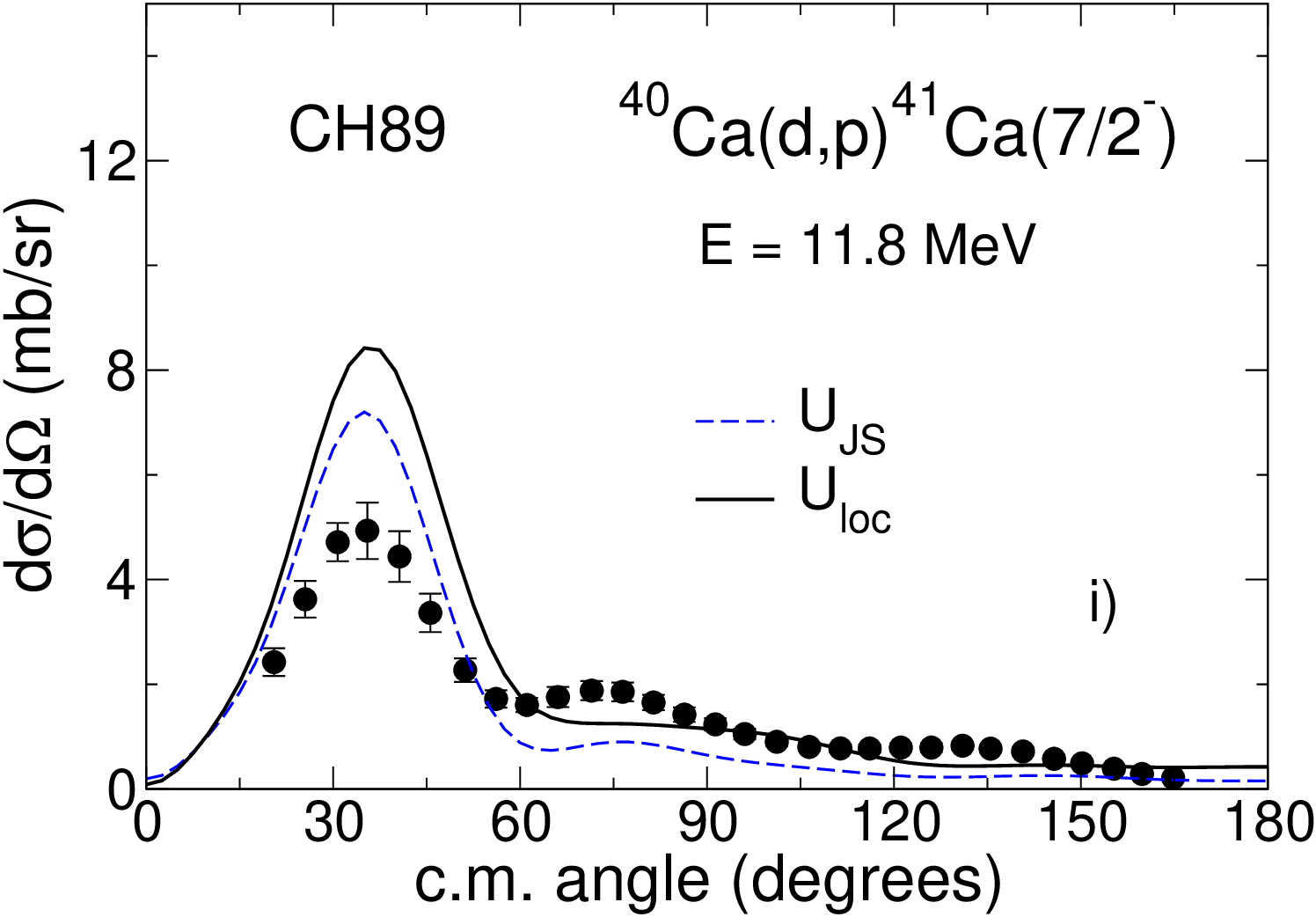}
 \caption{(Color online) 
 Angular distributions for the $^{16}$O$(d,p)^{17}$O, $^{36}$Ar$(d,p)^{37}$Ar and
 $^{40}$Ca$(d,p)^{41}$Ca reactions calculated with local adiabatic deuteron  potentials $U_{loc}$ obtained with
   energy-dependent GRZ ($a,b,c$),  energy-independent GR 
 ($d,e,f$) non-local optical nucleon potentials and energy-dependent local potential CH89 
 ($g,h,i$) in comparison  
 to the corresponding Johnson-Soper calculations.
 }\label{fig2}
\end{figure*}

According to the prescription of Sec. IV we solve the three-body Schr\"odinger equation
in the adiabatic approximation using energy-independent imaginary   potential (\ref{W})
evaluated at ${\cal E}_{\rm eff} = \frac{1}{2}E_d +\frac{1}{2}\la T_r\ra$. For $\frac{1}{2}\langle T_r\rangle$ we use the value of 57 MeV obtained
in Refs. \cite{Tim13a,Tim13b} for the Hult\'en $n-p$ potential. We also use
$\bar{V}_c = -1.08 +1.35Z/A^{1/3}$ MeV from \cite{Gia76,GRZ}. As explained in 
\cite{Tim13a,Tim13b}, in the adiabatic approximation the deuteron distorted wave
is found from a local-equivalent two-body model 
\beq
[E_d-T_R-U_c(R)-U_{loc}(R)]\,\chi^{(+)}_{\ve{k}_d}(\ve{R})=0,
\eeqn{locmod}
where $U_c(R)$ is the $d-A$ Coulomb potential and in the local-energy approximation the local potential $U_{loc}$  is a solution of the transcendental equation
\beq
U_{loc}(R) &=& M^{(0)}_0[U_{nA}({\cal E}_{\rm eff},R)+U_{pA}({\cal E}_{\rm eff},R)]
\eol &\times&
\exp\left[-\frac{\mu_d \beta_d^2}{2\hbar^2}(E_d - U_{loc}(R) - \bar{V}_c)\right].
\eeqn{Uloc}
In this equation $\mu_d$ is the deuteron-$A$ reduced mass and the constants $M^{(0)}_0$ and $\beta_d$   depend on $V_{np}$
and the nucleon non-locality range $\beta$  only. These constants are tabulated in \cite{Tim13b} for the
Hult\'en $n-p$ potential. The range $\beta_d$ has the meaning of an effective non-locality range
for the deuteron-target interaction and  is approximately equal to 0.46 which is close to $\beta/2$.
The moment $M^{(0)}_0$ is about 0.75 and its deviation from one determines the shift of
energy at which the local equivalents of ${\cal U}_{NA}({\cal E}_{\rm eff},R,R')$ should be evaluated
for use in Johnson-Tandy potential calculations. In this paper, we determine $U_{loc}$ by 
solving the transcendental equation (\ref{Uloc}) directly without constructing the local  
equivalent of ${\cal U}_{NA}({\cal E}_{\rm eff},R,R')$.

The local potentials $U_{loc}$ obtained for $d+^{16}$O, $d+^{36}$Ar and $d+^{40}$Ca
are shown in Fig. 1 ($a$-$c$). They are compared to the widely used
Johnson-Soper potentials $U_{JS}$ constructed
from the  energy-dependent local-equivalents of ${\cal U}_{NA}(E,R,R')$ 
taken at $E_d/2$:
\beq
 U_{JS}(R)&=& U_{loc}^n(E_d/2,R)+U_{loc}^p(E_d/2,R), \\  
 U_{loc}^n(E_d/2,R)&= &U_{nA}(E_d/2,R) 
\eol   &\times  &
\exp \left[ -\frac{\mu \beta^2}{2\hbar^2}\left(\frac{E_d}{2}-U_{loc}^n(E_d/2,R)\right)\right], \eol
 U_{loc}^p(E_d/2,R)&=&U_{pA}(E_d/2-\bar{V}_c,R) \eol
   \times 
&\exp & \left[ -\frac{\mu \beta^2}{2\hbar^2}\left(\frac{E_d}{2}-\bar{V}_c -U_{loc}^p(E_d/2,R)\right)\right].  \eol
\eeqn{PB}
One can see that the real parts of the deuteron local potentials $U_{loc}$ obtained
with GRZ  
 are shallower than the real parts of the corresponding Johnson-Soper potential
 $U_{JS}$. This is exactly what has been  seen already for the energy-independent non-local potential GR
 in \cite{Tim13a,Tim13b} and is shown again Fig. 1 ($d$-$f$). This reflects the fact that
 the real part of GRZ is energy-independent. The imaginary parts of   $U_{loc}$ obtained with GRZ are much deeper than those of $U_{JS}$, which is completely opposite to what happens
 for the imaginary parts of $U_{loc}$ and $U_{JS}$ obtained with the 
 energy-independent  potential GR. This happens because the 
 imaginary part of  GRZ  increases with $E$ and 
 at ${\cal E}_{\rm eff} = E_d/2+57$ MeV it takes on large values while the imaginary part
 of $U_{JS}$ is constructed from nucleon potentials taken at $4-8$ MeV where 
 $W_N(E)$ is small. On the contrary, for the energy-independent potential GR 
 the imaginary part of its local equivalent  slowly decreases with $E$. Taking this potential
 at $E_d/2+\Delta E$ leads to a smaller imaginary part. The different behaviour of the imaginary parts of $U_{loc}$ compared to that of $U_{JS}$ for GRZ and GR results in the difference manifestation of  the non-local effects
 in the corresponding $(d,p)$ differential cross sections.

To calculate the angular distributions of $(d,p)$ reactions we used the framework (\ref{exact})
in which the remnant term is absent and the $p-B$ channel is described by the $p-B$ distorted wave obtained with the $p-A$ optical potential thus neglecting recoil excitation and breakup effects (note that in Refs. \cite{Tim13a,Tim13b} the $p-B$ optical potential was used which means that the remnant term was neglected). 
These optical potentials were obtained from the local model corresponding to either the 
GRZ or the GR potential. The local proton scattering wave function was multiplied by a Perey factor \cite{Per63} with the proton non-locality range appropriate to GRZ or GR. 
No Perey effect was considered in the deuteron channel. Such effects would arise from 
the linear terms in the expansion of the nucleon optical potentials near $\ve{r}=0$ for which 
no averaging procedure has been developed. In the first place, these terms  will influence the shape of the
deuteron-target effective potential. The associated Perey effect will also  modify the
deuteron distorted wave, but only in the nuclear interior which does contribute significantly for
the $(d,p)$ considered here where peripheral contributions dominate.

Both the proton equivalent local potentials and the effective deuteron potentials $U_{loc}$
were read into the TWOFNR code \cite{2FNR} which was used to calculate
the $(d,p)$ cross sections in the zero-range approximation. The overlap functions for
$^{17}$O, $^{37}$Ar and $^{41}$Ca were represented by neutron single-particle wave functions
 obtained by fitting the Woods-Saxon potential
wells to reproduce their corresponding neutron separation energies. The standard radius $r_0 = 1.25 $ fm
and diffusenness $a = 0.65$ fm were used both for central and spin-orbit potentials while   the depth of the spin-orbit potential was 5 MeV. No Perey effect was used for the
neutron wave functions since we study only the relative change in the cross sections
caused by replacing the Johnson-Soper potentials by our new adiabatic distorting deuteron
potentials.

The $(d,p)$ cross sections calculated using   deuteron distorting adiabatic potentials
$U_{loc}(R)$
derived with GRZ and GR are shown in Fig. 2 ($a-c)$ and Fig. 2 $(d-f)$ respectively where they compared to the  calculations, performed with the Johnson-Soper potentials derived with GR and GRZ, and with experimental data. The $(d,p)$ cross sections are plotted with the spectroscopic factor $S = 1$ for $^{16}$O and $^{40}$Ca targets and  $S=0.5$ for $^{36}$Ar. One can see that the influence of non-locality for GRZ and GR is completely different. While for the energy-independent potential GR the cross sections increase and their shapes change insignificantly compared to Johnson-Soper calculations, for energy-dependent potential GRZ the cross sections decrease and the changes in the shapes of the angular distributions are more noticeable. This happens because the  Johnson-Soper potential corresponding to GRZ  has
much smaller imaginary part than that of $U_{loc}$ so that reduction of absorption leads to large cross sections. For the GR potential,  $U_{loc}$ has less absorption than $U_{JS}$ resulting in larger cross sections in comparison to the Johnson-Soper ones.  While the difference between the Johnson-Soper cross sections calculated with GRZ and GR can almost reach a factor of three (for $^{36}$Ar), the difference of the cross sections obtained in non-local model with GRZ and GR is much smaller, $\sim 20-30\%$ in the main peak. 

\subsection{Estimation of the deuteron distorting potential from phenomenological energy-dependent local nucleon optical potentials.}

Suppose we know some phenomenological energy-dependent optical potential
$U_{phen}(E,r)$. We assume that its underlying energy-dependent non-local potential
${\tilde U}_{NA}(E,r,r')$ has the Perey-Buck form
\beq
{\tilde U}_{NA}(E,r,r') = H(|\ve{r}-\ve{r}'|) U_{NA}(\frac{\ve{r}+\ve{r}'}{2},E),
\eeqn{}
where $H$ is given by Eq. (\ref{HPB})  with some  non-locality range $\beta$. We suppose
also that   $U_{NA}( E,r)$  is related to $U_{phen}(E,r)$ by the Perey-Buck transformations given by (\ref{PB}).
Then according to Sec. IV we need to use  formfactors
$ U_{NA}(E,r)$ taken at the energy $E = {\cal E}_{\rm eff} = E_d/2+ \la T_r \ra / 2$ treating them as energy-independent. The local $d-A$ distorting potential $U_{loc}$ is obtained from the transcendental equation (\ref{Uloc}) which can be rewritten as
\beq
U_{loc}(E_d,r) \exp\left[- \frac{\mu_d \beta_d^2}{2 \hbar^2} U_{loc}(E_d,r)\right] = 
{\cal V}(E_d,r),\,\,\,\,\,\,\,\,\,\,
\eeqn{Uloc2}
where 
\beq
{\cal V}(E_d,r) =  \exp\left[ \frac{\mu\beta^2}{2\hbar^2}{\cal E}_{\rm eff}- \frac{ \mu_d\beta_d^2}{2\hbar^2}(E_d - {\bar V}_c)\right]
\eol \times M_0\left[
U_{phen}^n({\cal E}_{\rm eff},r) \exp\left[ \frac{\mu \beta^2}{2\hbar^2}U_{phen}^n({\cal E}_{\rm eff},r)\right] \right.
\eol \left. +
U_{phen}^p({\cal E}_{\rm eff},r) \exp\left[ \frac{\mu \beta^2}{2\hbar^2}(U_{phen}^p({\cal E}_{\rm eff},r)-\bar{V}_c)\right] \right]. \eol
\eeqn{calV}
Thus, knowing phenomenological proton and neutron optical potentials we can calculate
${\cal V}(E_d,r)$ and  find $U_{loc}$ as a solution of Eq. (\ref{Uloc2}). In Fig.1 ($g-i$)
we have plotted
these solutions for $^{16}$O, $^{36}$Ar and $^{40}$Ca + d using the CH89 systematics for
$U_{phen}^n$ and $U_{phen}^p$  in comparison with the Johnson-Soper potential obtained also with CH89. We used $\beta = 0.85$ fm since the real part of the CH89 potential seems to be consistent with Perey-Buck non-local potential of this range.
The chosen value $\beta = 0.85$ fm corresponds to $\beta_d = 0.4$ fm and $M_0 = 0.78$ \cite{Tim13b}. We have also   solved Eq. (\ref{Uloc2}) by expanding the exponential
in its left-hand side up to second order terms but such solutions were not accurate enough.

Fig. 1 ($g-i$) shows that the distorting $d-A$ potential $U_{loc}$(CH89)  generated from an energy-dependent phenomenological local potential has a smaller real part
than $U_{JS}$(CH89), similar to the case of GR and GRZ. However, the imaginary part 
of $U_{JS}$(CH89) has more absorption in the internal nuclear region than has $U_{loc}$(CH89),  while 
being similar outside the nucleus. This is the consequence of the interplay
between the surface and volume absorption in CH89 and the fact that the imaginary part has to be evaluated at higher energies where the volume absorption grows. Nevertheless, the differential cross sections, plotted in Fig.2 ($g-i$) show a similar influence of non-locality to what has been observed with energy-independent potential GR. The cross sections increase in the main
peak by 16$\%$, 10$\%$ and 18$\%$ for the $^{16}$O, $^{36}$Ar and $^{40}$Ca targets respectively. In comparison,  an increase of 15$\%$ is obtained with GR. The absolute cross sections in the area of the main experimental peak
obtained in non-local calculations with three different nucleon optical potentials
differ by 20$\%$  in the case of $^{16}$O and $^{36}$Ar and by 30$\%$ in the case of $^{40}$Ca. This is comparable to  the dependence on optical potential parameters found typically in  distorted-wave Born approximation calculations.

\section{Discussion and Conclusions}

We have suggested an approximate practical way of dealing with explicitly energy-dependent non-local nucleon optical potentials when calculating the $(d,p)$ reactions within the $A + n + p$ model space. We have shown that within an approximation consistent with the adiabatic model of $(d,p)$ reactions the problem of using energy dependent nonlocal nucleon potentials is reduced to the problem of calculating the $(d,p)$ cross sections with energy-independent nonlocal potentials, the solution of which has been found in \cite{Tim13a,Tim13b}. 
It is important to note that in obtaining our results we have assumed that any explicit energy dependence of optical potentials arises from many-body effects. In our treatment we start from an approximate three-body Hamiltonian with non-local nucleon optical potential operators at an energy fixed by the incident deuteron energy. Our prescription is based on the identification of the particular feature  of the $(d,p)$ amplitude (small $n-p$ separations) that is relevant to the calculation of a particular form of the many-body $(d,p)$ transition operator. It is not a prescription that is intended to be useful for the whole of configuration space. For example, it can not be used in calculations of $d-A$ elastic scattering.  

It is also important to note that in deriving the three-body effective Hamiltonian
we have ignored the $U_{pA}\frac{Q_A}{e}U_{nA}+.....$ terms in Eq. (\ref{Umult}). These terms reflect the fact that the many-body problem asociated with the $n+p+A$ system cannot be mapped exactly onto a three-body model with Hamiltonian $T_3+V_{np}+U_p+U_n$. The neglected terms are non-local energy dependent three-body forces that are a necessary consequence of the underlying many-body problem. They describe physical processes in which the target is excited by the incident neutron and de-excited by the proton and so on in higher orders. Terms in which successive target excitations are caused by the same nucleon are taken into account by the $U_n$ and $U_p$ operators. The quantitative significance of the $U_{pA}\frac{Q_A}{e}U_{nA}+.....$  terms is unknown.

 The phenomenological optical potentials can be assumed  to take  antisymmetrization effects into account.
These effects are believed to make important contributions to non-locality and are taken into account by the treatment of Timofeyuk and Johnson in \cite{Tim13a,Tim13b}. However, for the $d+A$ system there are additional antisymmetrization effects that cannot be taken into account in this way. For a discussion  of their quantitative significance see Tostevin, Lopes and Johnson \cite{TLJ87}. 

The prescription of the present paper is simple enough to be widely used in analysis of   $(d,p)$ experiments. However, it requires knowledge of the  
energy-dependence of non-local optical potentials. At present, we are not aware of any potential of such a kind \cite{Mah13} except for the GRZ potential. While this potentials generates reasonable $d-A$ effective local potentials it clearly gives unphysical local Johnson-Soper potentials that strongly overestimate  $(d,p)$ cross sections 
at the low incident deuteron energies frequently used at several modern radioactive beam facilities. This shows that more studies, using both phenomenological and microscopic models, are needed to pin down the energy-dependence of nucleon optical potentials. 
 
We have also suggested a simple way of correcting the $d-A$ potentials for non-locality when the
energy-dependence of non-local  nucleon optical  potentials is unknown but energy-dependent local systematics are available. We have shown that if a specific assumption  about the form of energy-dependent non-local potentials is made, then the local distorting deuteron potentials for adiabatic $(d,p)$ calculations are obtained as solutions of a transcendental equation. While we believe that this assumption should be valid for the targets considered here, in general it may not be correct. For example, according to GRZ, the asymmetry term of the optical potential has much wider non-locality. Therefore, we can expect that for $Z\neq N$ targets Eqs. (\ref{Uloc}) and (\ref{calV})  could be modified.

Finally, we did not intend to provide new values for the spectroscopic factors and asymptotic normalization coefficents obtained from the analysis of $(d,p)$ reactions using our new prescription. Our aim here is to  clarify  how the explicit energy-dependence of non-local optical potentials affects known results of adiabatic $(d,p)$ calculations. We do have  preliminary ideas  of what changes in spectroscopic factors and asymptotic normalization coefficients could be expected, but the details of any   change will depend on how well we understand the energy dependence of non-local optical  potentials. On the other hand, our study is not comprehensive and many issues remain outstanding. These include 
the influence of the deuteron $d$-state on the effective $d-A$ potential, corrections to the approximate treatment of non-locality in \cite{Tim13a},\cite{Tim13b},  non-adiabatic corrections,  additional antisymmetrization effects, multiple scattering etc. These questions present a challenge to the development of the $(d,p)$ reaction theory and answering them is important for the correct
 interpretation of  measured $(d,p)$ cross
sections in terms of the nuclear structure quantities.

\section{Acknowledgement}

We greatfully acknowledge  the support from
the UK STFC ST/J000051/1 grant.

\section*{Appendix}

\subsection{The operator $U$.}
Here, we derive the formula (\ref{U}) from Section II for the operator $U$ whose matrix elements can be used to give formally exact expressions for projections of the many-body scattering state $\mid \Psi^{(+)}_{\ve{k}_d,A} \ra $ corresponding to a deuteron of momentum $\ve{k}_d$ in its ground state $\mid \phi_0 \ra $ incident on a target in its ground state $\mid \phi_A \ra $, which we will assume has spin zero for simplicity here. The relevant projection operators $P_A$ and $Q_A$ project onto the ground and excited states of nucleus $A$ respectively and satisfy
\beq
P_A+Q_A=1, \,\,\,\,\,\, P_A\mid \phi_A \ra = \mid \phi_A \ra, \,\,\,\,\,\, Q_A\mid \phi_A \ra = 0, \eol 
P_A^2=P_A, \,\,\,\,\,\,\,\,\,\,\,\,\,\,\,\,  Q_A^2=Q_A, \,\,\,\,\,\,\,\,\,\,\,\, P_AQ_A = 0. \,\,\,\,\,\, \,\,\,\,\,\, 
\label{PAQA}\eeq
The state $\mid \Psi^{(+)}_{\ve{k}_d,A} \ra $ is obtained by taking the limit $\epsilon \rightarrow +0$ of the state $\mid \Psi^{(\epsilon)}_{\ve{k}_d,A} \ra $  that satisfies the equation
 \begin{eqnarray}
 (E+\imath \epsilon -H)\mid \Psi^{(\epsilon)}_{\ve{k}_d,A} \ra =\imath \epsilon \mid \ve{k}_d,\phi_0\phi_A \ra ,
 \label{ExactEq}\end{eqnarray}
where $H$ is the  Hamiltonian and $E$ is the total energy of the $A+n+p$, which in the notations of Section II is
\beq H=T_3+V_{np}+U^0_n(n)+U^0_p(p)+H_A+\Delta v_{nA}+\Delta v_{pA}. \eol  \label{H}  
\eeq  
For $\epsilon \neq 0$ the function $\mid\Psi^{(\epsilon)}_{\ve{k}_d,A}\ra $ is uniquely defined by the inhomogeneous Eq. (\ref{ExactEq}). No further statements of boundary conditions are required. It follows from Eq. (\ref{ExactEq}) that  $\mid\Psi^{(\epsilon)}_{\ve{k}_d,A}\ra - \mid \ve{k}_d, \phi_o \phi_A \ra$ is square-integrable. As discussed in Goldberger and Watson, Ref. \cite{goldberger}, Ch.5,``Formal Scattering Theory'', the function $\mid\Psi^{(\epsilon)}_{\ve{k}_d,A}\ra $  is used as a stepping stone to the calculation of transition matrices for all final channels using appropriate expressions in which the limit $\epsilon \rightarrow  +0$ can be taken without ambiguity. They show that their methods can be justified by wave packet arguments.

  
Following Feshbach \cite{Fes58} we obtain equations coupling the $P_A$ and $Q_A$ components of $\mid \Psi^{(\epsilon)}_{\ve{k}_d,A} \ra$, defined as
\beq
\Psi_P   = P_A\mid \Psi^{(\epsilon)}_{\ve{k}_d,A} \ra, \,\,\,\,\,\,\,\,\,\,\,
\Psi_Q = Q_A\mid \Psi^{(\epsilon)}_{\ve{k}_d,A} \ra\,\,\,\,\,
\eeqn{}
 by operating with $P_A$ and $Q_A$ in turn on both sides of Eq. (\ref{ExactEq}) to obtain
 \begin{eqnarray}
 (E+\imath \epsilon -P_AHP_A)\Psi_P-P_AHQ_A\Psi_Q &=&\imath \epsilon \mid \ve{k}_d,\phi_0\phi_A \ra , \eol
 (E+\imath \epsilon -Q_AHQ_A)\Psi_Q-Q_AHP_A\Psi_P & =& 0. \label{ExactEq2} \end{eqnarray}
From the second equation we deduce
\begin{eqnarray}
 \Psi_Q=\frac{1}{ (E+\imath \epsilon -Q_AHQ_A)}Q_AHP_A\Psi_P, \label{ExactEq3}\end{eqnarray}
 and substituting this result into the first of Eq. (\ref{ExactEq2}) we get an equation for $\Psi_P$ alone:
 \begin{eqnarray}
 (E+\imath \epsilon -P_AHP_A)\Psi_P &-&P_AHQ_A\eol
&\times & \frac{1}{ (E+\imath \epsilon -Q_AHQ_A)}Q_AHP_A\Psi_P\eol &=&\imath \epsilon \mid \ve{k}_d,\phi_0\phi_A \ra, \eol \label{ExactEq4}\end{eqnarray}
 where $\mid \ve{k}_d,\phi_0\phi_A \ra$ describes a plane wave deuteron in its ground state incident on $A$ in its ground state  $\mid \phi_A \ra$.

  For a spin-zero target $A$ the projection $\Psi_P$  will have the form $\mid\Psi^{(\epsilon)}_{\ve{k}_d}(n,p)\ra \mid \phi_A \ra$. We obtain an equation for the function of $n$ and $p$ coordinates, $\mid\Psi^{(\epsilon)}_{\ve{k}_d}(n,p)\ra$, by taking the inner product of Eq. (\ref{ExactEq4})  with $\mid \phi_A \ra$. Using the notation used in
  Eqs. (\ref{U}) and (\ref{e}) we get
 \begin{eqnarray}
 (E_3+\imath \epsilon -T_3-V_{np}-U_n^0(n)-U_p^0(p)  \,\,\,\,\,\,\,\,\,\,\,\,\,\,\,\,\,\,\,\, \,\,\,\,\,\,\,\,\,\,\,\,\,\,\,\,\,\,\,\,\,\,\,\,\,\,\, \eol
 -\la \phi_A \mid (\Delta v_{nA}+\Delta_{pA})\mid \phi_A \ra &
 \eol
 -\la \phi_A \mid(\Delta v_{nA}+\Delta_{pA})Q_A\frac{1}{ e -Q_A(\Delta v_{nA}+\Delta_{pA})Q_A}
 \eol \times
 Q_A(\Delta v_{nA}+\Delta_{pA})\mid \phi_A \ra ) \mid \Psi^{(\epsilon)}_{\ve{k}_d}(n,p) \ra 
 =\imath \epsilon \mid \ve{k}_d,\phi_0 \ra , \eol \label{ExactEq5}\end{eqnarray}
where $E_3 = E-E_A$ and we have used the fact that $E+\imath \epsilon -T_3-V_{np}-U_n^0(n)-U_p^0(p)$ commutes with $P_A$ and $Q_A$ so that
\beq 
P_A(E+\imath \epsilon -T_3-V_{np}-U_n^0(n)-U_p^0(p))P_A
\eol = (E+\imath \epsilon -T_3-V_{np}-U_n^0(n)-U_p^0(p))P_A, \eol
Q_A(E+\imath \epsilon -T_3-V_{np}-U_n^0(n)-U_p^0(p))Q_A
\eol = (E+\imath \epsilon -T_3-V_{np}-U_n^0(n)-U_p^0(p))Q_A, \eol
 P_A(E+\imath \epsilon -T_3-V_{np}-U_n^0(n)-U_p^0(p))Q_A&=&0. \eol \label{PAePA}\eeq
 We have also used the fact that
because the operator 1/(denominator) in Eq. (\ref{ExactEq5}) appears with $Q_A$ on either side  then (i) only excited state eigenvalues of $H_A$ ever appear and  (ii) $H_A$  is diagonal in the basis of states of $A$. Therefore, we can replace $Q_A(H_A-E_A)Q_A$  by $H_A-E_A$.
 
 It is convenient here to define an operator $U$ in the complete $n+p+A$ many-body space by the equation
 \begin{eqnarray}
 U=  (\Delta v_{nA}+\Delta_{pA})+(\Delta v_{nA}+\Delta_{pA})Q_A \eol \times
 \frac{1}{ e -Q_A(\Delta v_{nA}+\Delta_{pA})Q_A}
 Q_A(\Delta v_{nA}+\Delta_{pA}).\eol \label{defU}\end{eqnarray}
 In terms of $U$ the  Eq. (\ref{ExactEq5}) can be written as
 \begin{eqnarray}
 (E_3+\imath \epsilon - T_3-V_{np}-U_n^0(n)-U_p^0(p) \,\,\,\,\,\,\,\,\,\,\, \eol
 -\la \phi_A \mid U\mid \phi_A \ra ) \mid \Psi^{(\epsilon)}_{\ve{k}_d}(n,p) \ra\
 =\imath \epsilon \mid \ve{k}_d,\phi_0 \ra , \label{ExactEq6}\end{eqnarray}
 We see that the wavefunction $\mid \ve{k}_d,\phi_0 \ra$ of the three-body model is determined by the target ground state matrix element of $U$, $\la \phi_A \mid U\mid \phi_A \ra$.
 
  We use the identity
 \begin{eqnarray}
 \frac{1}{ e -O}&=&
 \frac{1}{e }
 +\frac{1}{ e }O\frac{1}{ e -O},  \label{Gid}\end{eqnarray}
 where $O$ is an arbitrary operator. Identifying $O$ with $Q_A\sum_N\Delta v_{NA}Q_A$ we can show that $U$ satisfies the equation (\ref{U}) of Section II.

 \subsection{Derivation of a multiple scattering expansion for $U$.}
The operator $U$ is the unique solution of the equation
\begin{eqnarray}
U=\Delta v_{nA}+\Delta v_{pA} 
+(\Delta v_{nA}+\Delta v_{pA})\frac{Q_A}{e}U. \,\,\,\,\,\,\,\,\,\,
\label{2U}
\end{eqnarray}
We derive the multiple scattering expansion of $U$ using the techniques of Watson \cite{goldberger}. We  introduce two new operators $W_n$ and $W_p$ defined  in terms of $U$ by
\begin{eqnarray}
W_N=\Delta v_{NA} 
+\Delta v_{NA}\frac{Q_A}{e}U, \,\,\,N=n,p.
 \label{WN}
\end{eqnarray}
By adding the formulae for $N=n$ and $N=p$ it can be seen that $W_n +W_p$ satisfies Eq. (\ref{2U}) and therefore
\begin{eqnarray}
U=W_n +W_p.\label{WNU}
\end{eqnarray}
We next show how these operators are related to  the operators $U_{NA}$ used in the text and defined by
\begin{eqnarray}
U_{NA}= \Delta v_{NA} +\Delta  v_{NA}\frac{Q_A}{e}U_{NA}. 
\label{2UNA}
\end{eqnarray}
A useful way of expressing the solution of this equation is to take the second term on the right over to the left and deduce
\begin{eqnarray}
U_{NA}=\frac{1}{(1-\Delta  v_{NA}\frac{Q_A}{e}) } \Delta  v_{NA}.  
 \label{3UNA}
\end{eqnarray}
Performing the analogous step in Eq. (\ref{WN}) with $N=n,p$ in turn and with $U$ replaced by $W_n +W_p$ we obtain
\beq W_n &=&U_{nA} 
+U_{nA}\frac{Q_A}{e}W_p, 
\eol
W_p &=&U_{pA} 
+U_{pA}\frac{Q_A}{e}W_n. \label{WNUN}\eeq
 Substituting the $W_p$ formula into the right-hand-side of the $W_n$ equation and the $W_n$ formula into the right-hand-side of the $W_p$  we obtain the uncoupled equations
\beq W_n&=&U_{nA} 
+U_{nA}\frac{Q_A}{e}U_{pA}+U_{nA}\frac{Q_A}{e}U_{pA}\frac{Q_A}{e}W_n 
\eol
W_p&=&U_{pA} 
+U_{pA}\frac{Q_A}{e}U_{nA}+U_{pA}\frac{Q_A}{e}U_{nA}\frac{Q_A}{e}W_p.\label{WNUNuncoupled}\eeq
In the standard way we iterate both of these equations and then construct $U=W_n+W_p$. The first few terms give the multiple scattering series discussed in Section III.

\subsection{Derivation of the inversion formula, Eq. (\ref{oneovere})}
In order to understand the origin of the approximate inversion of the operator $(e-Q_A\Delta v_{NA}Q_A)$ given in Eq. (\ref{oneovere}) it is first necessary to look at the ADWA from a different point of view from that customarily used. 

In section \ref{Section II},  Eq. (\ref{exact}) we introduced $\Psi^{(+)}_{\ve{k}_d}(n,p)$, the projection of the full $d+A$ many body wavefunction on to the ground state of $A$. It is the limit $\epsilon \rightarrow +0$ of the function $\Psi^{(\epsilon)}_{\ve{k}_d}(n,p)$ that satisfies
\begin{eqnarray}
 (E_3 + \imath \epsilon -H_{\rm eff})\Psi^{(\epsilon)}_{\ve{k}_d}(n,p)=\imath \epsilon \mid \phi_0 \ve{k}_d \rangle,  \label{Psiepsilon}   
\end{eqnarray}
where $H_{\rm eff}$ is the effective three-body Hamiltonian,
 \begin{eqnarray}
 H_{\rm eff}=T_R+H_{np}+ {\cal U}(n,p),  \label{Heff1}   
\end{eqnarray}
with $H_{np}=T_{np}+V_{np}$  and some arbitrary interaction ${\cal U}(n,p)$ which can be an operator in the $\ve{r}$ and  $\ve{R}$ degrees of freedom. A formal solution to Eq. (\ref{Psiepsilon}) is
\begin{eqnarray}
 \Psi^{(\epsilon)}_{\ve{k}_d}(n,p)=\frac{\imath \epsilon}{(E_3 + \imath \epsilon -H_{\rm eff})} \mid \phi_0 \ve{k}_d \rangle.  \label{Psiepsilonsol}   
\end{eqnarray}
From this point of view we see that the solution of the three-body problem for $\Psi^{(\epsilon)}_{\ve{k}_d}(n,p)$ is equivalent to inverting the operator $(E_3 + \imath \epsilon -H_{\rm eff})$. We now show that the ADWA can be regarded as an approximate inversion of this operator within the space of Weinberg states.

The Weinberg basis $\{\phi^W_i\}$ associated with $H_{np}$ was introduced by Johnson and Tandy \cite{JT} as being well adapted to the evaluation of the components of $\Psi^{(\epsilon)}_{\ve{k}_d}(n,p)$ that dominate the $A(d,p)B$ matrix element, Eq. (\ref{exact}) \cite{PTJT}. 
The Weinberg states $\phi^W_i$ satisfy the orthonormality relation
 \beq
\la \phi^W_i\mid V_{np} \mid \phi^W_k\ra=- \delta_{ik}.
\label{orth}\eeq
and the components $\mid \chi_{i}^{(\epsilon)}\ra$ of $\mid \Psi_{\ve{k}_d}^{(\epsilon)}\ra$ in this basis are    
\beq
\mid \chi_{i}^{(\epsilon)}\ra=
-\la \phi^W_i\mid V_{np} \mid \Psi_{\ve{k}_d}^{(\epsilon)}\ra.
\label{chi}\eeq
The ADWA is obtained by multiplying Eq. (\ref{Psiepsilon}) by $-\la \phi^W_1\mid V_{np}$ on the left, integrating
over $\ve{r}$ and neglecting all couplings between   $\mid \chi_{1}^{(\epsilon)}\ra$ and
all other $i\neq 1$ Weinberg components. This results in 
\beq
 -\la \phi^W_1\mid V_{np} (E_3 + \imath \epsilon -H_{\rm eff})\mid \phi^W_1\ra 
 \mid \chi_{1}^{(\epsilon)}\ra = \imath \epsilon C_1^{-1} |\ve{k}_d\ra, \eol
\eeqn{f1}
where $C_1$ is  the constant relating the first Weinberg state $\phi^W_1$ to the deuteron wave function $\phi_0$ by $\phi^W_1 = C_1 \phi_0$  and satisfies $\mid C_1\mid ^2 \la \phi_0\mid V_{np} \mid \phi_0\ra = -1$ \cite{JT}.
On the other hand, after multiplying Eq. (\ref{Psiepsilonsol})  by $-\phi^W_1V_{np}$ on the left and integrating over $\ve{r}$ we get:
\beq
\mid \chi_{1}^{(\epsilon)}\ra = -i \epsilon C_1^{-1}
\la \phi^W_1\mid V_{np} \frac{1}{E_3 + \imath \epsilon -H_{\rm eff}}\mid \phi^W_1\ra 
 \mid\ve{k}_d\ra . \eol
\eeqn{f2}
Combining  Eqs. (\ref{f2}) into (\ref{f1}) we get
\beq
& \imath \epsilon  &\la   \phi^W_1\mid V_{np}(E_3 + \imath \epsilon -H_{\rm eff})\mid \phi^W_1\ra  \eol
& \times  &
 \la \phi^W_1\mid V_{np} \frac{1}{E_3 + \imath \epsilon -H_{\rm eff}}\mid \phi^W_1\ra 
 \mid\ve{k}_d\ra = \imath \epsilon \mid\ve{k}_d\ra,  \,\,\,\,\,\,\,\,\,\,\,\,
\eeqn{f3}
which is satisfied if
\beq
\la \phi^W_1\mid V_{np} \frac{1}{E_3 + \imath \epsilon -H_{\rm eff}}\mid \phi_1^W\ra  \eol
  = \frac{1}{\la \phi^W_1\mid V_{np} (E_3 + \imath \epsilon -H_{\rm eff})\mid \phi_1^W\ra}.
\eeqn{f4}
Taking Eq. (\ref{phi1}) and $ \phi^W_1 = C_1 \phi_0$ into account in Eq. (\ref{f4}), we see that the approximation  (\ref{oneovere}) 
follows from the same approximate inversion idea as used in the ADWA.

We now show how this approach can be made the basis for making systematic improvements to the ADWA. In the Weinberg basis an arbitary operator $\hat{O}$ is represented by the matrix
\beq
\hat{O}^W_{ij}=-\la \phi_i\mid V_{np} \hat{O}\mid \phi_j\ra.
\label{W4}\eeq
Eq.(\ref{Psiepsilonsol}) tells us that these components are given  by
\beq
\mid \chi_{i}^{(\epsilon)}\ra=-\sum_k\imath \epsilon G^W_{ik}
\la \phi^W_k\mid  V_{np} \mid \phi_0 \ve{k}_d \ra,
\label{LSWein}\eeq
where
\beq
G^W_{ik} =-\la \phi^W_i\mid  V_{np}\frac{1}{E_3+\imath \epsilon-H_{{\rm eff}}(n,p)} \mid \phi^W_k \ra. 
\label{GW}\eeq
Note that each matrix element $G^W_{ik}$ is an operator in the space of  $\ve{R}$.
Using the relation  $\mid \phi^W_1 \ra=C_1\mid \phi_0 \ra,$ we have
\beq
- \la \phi^W_k\mid V_{np} \mid \phi_0 \ve{k}_d \ra =\delta_{k1}\frac{1}{C_1}\mid \ve{k}_d \ra,
\label{k0}\eeq
and Eq. (\ref{LSWein}) reduces to
\beq
\mid \chi_{i}^{(\epsilon)}\ra= \frac{\imath\epsilon}{C_1} G^W_{i1} \mid \ve{k}_d \ra. 
\label{LSWein3}\eeq

It was shown in \cite{PTJT} that for $(d,p)$ reactions  in a range of incident energies of current interest the $(d,p)$ transition matrix, Eq. (\ref{exact}), is dominated by the first Weinberg component $\mid \chi_{1}^{(\epsilon)}\ra$. According to Eq. (\ref{LSWein3}) an exact expression for this is
\beq
\mid \chi_{1}^{(\epsilon)}\ra= \frac{\imath\epsilon}{C_1} G^W_{11} \mid \ve{k}_d \ra, 
\label{LSWein4}\eeq
which only involves the single matrix element $G^W_{11}$.

The evaluation of $G^W_{ik}$ requires the inversion of the matrix
\beq
D^W_{lm} =-\la \phi^W_l\mid  V_{np}(E_3+\imath \epsilon-H_{\rm eff})) \mid \phi^W_m \ra ,
\label{DW}\eeq 
so that 
\beq D^WG^W=1. \label{DWGW}\eeq
By writing $D^W_{ij}=D^W_{ii}\delta_{ij}+D^W_{ij}(1-\delta_{ij})$ it is straightforward to show that an exact alternative to the condition (\ref{DWGW}) is
\beq G^W=\frac{1}{D^{W\mathrm{diag}}}+\frac{1}{D^{W\mathrm{diag}}}D^{W\mathrm{nondiag}}G^W, \label{DWGW2}\eeq
where
\beq D^{W\mathrm{diag}}_{ij}&=&D^W_{ii}\delta_{ij},\nonumber \\D^{W\mathrm{nondiag}}_{ij}&=&D^W_{ij}(1-\delta_{ij}). \label{Ddiag}\eeq
In particular
\beq G^W_{11}=\frac{1}{D^W_{11}}+\frac{1}{D^W_{11}}\sum_{j\neq 1}D^W_{1j}G^W_{j1} \label{DWGW3}\eeq 

The essence of the ADWA is that it  ignores all couplings between the first Weinberg component and all others \cite{JT}. In this limit we deduce from Eq. (\ref{DWGW3}) that
\beq G^W_{11}&=&\frac{1}{D^W_{11}}\nonumber \\
&=&-\frac{1}{\la \phi^W_1\mid  V_{np}(E_3+\imath \epsilon-H_{\rm eff}) \mid \phi^W_1 \ra}.
\label{GW112}\eeq
in agreement with Eq. (\ref{f4}).
The error is of second order in the off-diagonal elements $\la \phi^W_1\mid V_{np}(E_3+\imath \epsilon-T_R-H_{\rm eff}(n,p))\mid \phi^W_i\ra $ with $ i\neq 1$. 

This approach can be developed by iteration of Eq. (\ref{DWGW2}) to give corrections to the ADWA.

\subsection{Averaging $T_p+V_{np}$ and $T_n+V_{np}$}

In Section IV we have shown that
 the expression for the ADWA distorting potential requires the evaluation of matrix element  (\ref{oneovere}).  The denominator in the neutron contribution contains the operator $T_n$, as appears in the expression for neutron optical potential, and the operator $T_p+V_{np}$  for which we want to deduce an average value consistent with ADWA. Here  we evaluate the average $\langle \phi_1\mid T_p+V_{np}\mid \phi_0\rangle$ acting on $\chi^{(+)}_{\ve{k}_d}(\ve{R})$.  
Since
\beq
T_p = \frac{1}{2}(T_R+T_r)-\frac{\hbar^2}{2M_N}\ve{\nabla}_r.\ve{\nabla}_R,
\eeqn{}
we obtain
\begin{eqnarray}
 \langle \phi_1\mid T_p\mid \phi_0\rangle \,\chi^{(+)}_{\ve{k}_d}(\ve{R})
=\frac{1}{2}(T_R+\la T_r\ra)\,\chi^{(+)}_{\ve{k}_d}(\ve{R})
\eol
-\frac{\hbar^2}{2M_N}\left[\int d\ve{r}\phi_1^*( \ve{r})\ve{\nabla}_r\phi_0(\ve{r})\right].\ve{\nabla}_R\chi^{(+)}_{\ve{k}_d}(\ve{R})\nonumber \\
=\frac{1}{2}(T_R+\la T_r\ra)\,\chi^{(+)}_{\ve{k}_d}(\ve{R})
, \label{phi1phi0chiTp}\end{eqnarray}
  where the last line contains the quantity  
  \beq \la T_r\ra=\langle \phi_1\mid T_r\mid \phi_0\rangle  \label{Tr}\eeq
introduced by Timofeyuk and Johnson in \cite{Tim13a}. In obtaining the last line in Eq. (\ref{phi1phi0chiTp}) we have used that fact that $\phi_0$  and   $\phi_1$ have the same parity  and that $\langle \phi_1\mid \phi_0\rangle=1$. 
Using  the exact result 
\begin{eqnarray}   
\la \phi_1 | V_{np} |\phi_0 \ra =-\epsilon_0-\langle T_r \rangle.
 \label{phi1Vnpphi0}\end{eqnarray}
we obtain
 \beq
 \langle \phi_1\mid T_p+V_{np}\mid \phi_0\rangle \chi^{(+)}_{\ve{k}_d}(\ve{R}) =
[\frac{1}{2}(T_R - \la T_r\ra ) - \epsilon_0] \chi^{(+)}_{\ve{k}_d}(\ve{R}). \eol
 \eeqn{}
 The same results is  obtained for the average $\langle \phi_1\mid T_n+V_{np}\mid \phi_0\rangle $
 that is relevant to the calculation of the proton contribution to the  distorting potential in the deuteron channel.

\end{document}